\documentclass[10pt,journal,compsoc]{IEEEtran}

\usepackage{mathptmx}
\usepackage{fancyhdr}
\usepackage[normalem]{ulem}
\usepackage[hyphens]{url}
\usepackage{microtype}
\usepackage{xcolor, soul}
\usepackage{subcaption}
\usepackage{siunitx}
\usepackage{graphicx}
\usepackage{booktabs} 
\usepackage{balance}
\usepackage{tabulary}
\usepackage{amsmath, amssymb}
\usepackage{xspace}
\usepackage{caption}
\usepackage{multirow}
\newcommand{\ignore}[1]{}
\colorlet{lighterblue}{blue!15}
\colorlet{lightergreen}{green!20}

\newcommand{\CAP}{\emph{CAP}\xspace}
\newcommand{\NSP}{\emph{NSP}\xspace}

\ifCLASSOPTIONcompsoc
  \usepackage[nocompress]{cite}
\else
  \usepackage{cite}
\fi

\usepackage[bookmarks=true,breaklinks=true,letterpaper=true,colorlinks=true,linkcolor=black,citecolor=blue,urlcolor=black, draft=true]{hyperref}

\begin{document}
\title{Stagioni: Temperature management to enable near-sensor processing for energy-efficient high-fidelity imaging}

\author{Venkatesh Kodukula,~\IEEEmembership{Member,~IEEE,}
        Saad Katrawala,
        Britton Jones,
        Carole-Jean Wu,~\IEEEmembership{Senior Member,~IEEE,}
        and Robert LiKamWa
\IEEEcompsocitemizethanks{\IEEEcompsocthanksitem V. Kodukula, S. Katrawala, B. Jones, and R. LiKamWa are with the School
of Electrical, Computer, and Energy Engineering, and, C.-J. Wu is with the School of Computing, Informatics, and Decision Systems Engineering, Arizona State University, Tempe, AZ, 85281.\protect\\
E-mail: \{vkoduku1, skatrawa, bsjones7, carole-jean.wu, likamwa\}@asu.edu
}
}

%

\IEEEtitleabstractindextext{%
\begin{abstract}
Vision processing on traditional architectures is inefficient due to energy-expensive off-chip data movement.
Many researchers advocate pushing processing close to the sensor to substantially reduce data movement.
However, continuous near-sensor processing raises the sensor temperature, impairing the fidelity of imaging/vision tasks.
We characterize the thermal implications of using 3D stacked image sensors with near-sensor vision processing units.
Our characterization reveals that near-sensor processing reduces system power but degrades image quality.
For reasonable image fidelity, the sensor temperature needs to stay below a threshold, situationally determined by application needs.
Fortunately, our characterization also identifies opportunities -- unique to the needs of near-sensor processing -- to regulate temperature based on dynamic visual task requirements and rapidly increase capture quality on demand.
Based on our characterization, we propose and investigate two thermal management strategies -- stop-capture-go and seasonal migration -- for imaging-aware thermal management.
We present parameters that govern the policy decisions and explore the trade-offs between system power and policy overhead.
Our evaluation shows that our novel dynamic thermal management strategies can unlock the energy-efficiency potential of near-sensor processing.
For our evaluated tasks, our strategies save up to 53\% of system power with negligible performance impact and sustained image fidelity.
\end{abstract}

}

\maketitle


\section{Introduction}

\begin{figure}
    \centering
    \begin{subfigure}[b]{\columnwidth}
        \centering \captionsetup{width=.7\linewidth}%
        \includegraphics[width=.7\textwidth]{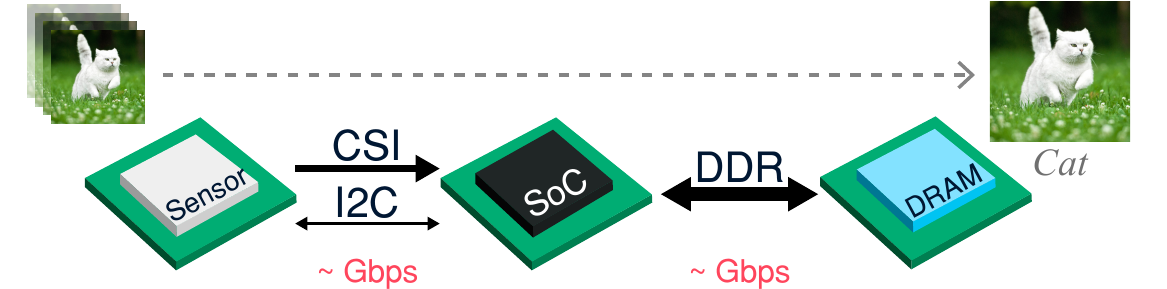}
        \caption{Traditional vision pipeline.}
        \label{fig:trad_vision_pipeline}
        \vspace{1em}
    \end{subfigure}
    \begin{subfigure}[b]{\columnwidth}
     \centering\captionsetup{width=.6\linewidth}%
         \includegraphics[width=.7\textwidth]{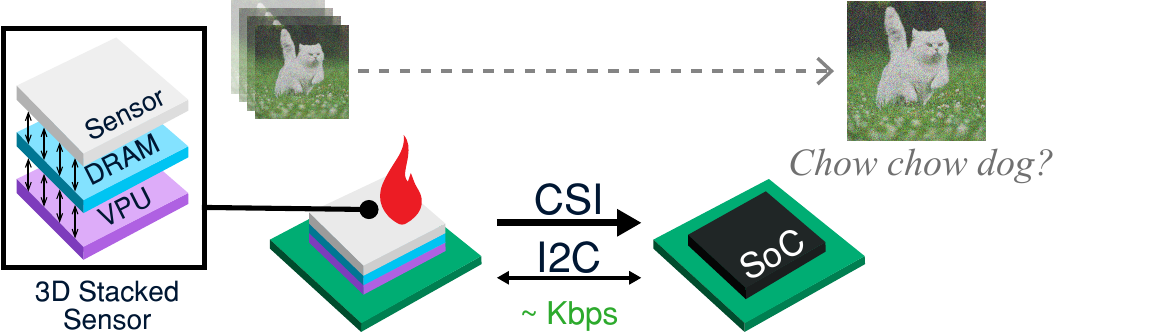}
         \caption{Near-sensor vision pipeline.}
        \label{fig:near_sens_vision_pipeline}
    \end{subfigure}
\caption{Due to energy-expensive interface data movements, traditional pipelines are inefficient. Near-sensor processing helps greatly reduce data traffic promoting energy-efficiency. However, it generates heat increasing sensor temperature, thereby resulting in noisy images potentially degrading task accuracy.}
\vspace{-1.5em}

\end{figure}

Imaging and vision systems allow computing systems to sense real-world visual situations and to capture images for human consumption.
Camera-enabled devices can now perform a wide range of visual tasks such as detecting and tracking objects~\cite{FB_AR}, constructing spatial maps for augmented reality~\cite{FB_AR2,hololens}, and providing driverless navigation assistance~\cite{nvidia_self_driving}.
Unfortunately, imaging requires high data rates to transfer pixel data from the image sensor to computational units.
In traditional systems (Fig.~\ref{fig:trad_vision_pipeline}), where the computational units are separated from the sensor via long interconnects, e.g., ribbon cables, these data rates create bottlenecks to energy efficiency and processing.
Thus, current vision systems result in power profiles on the order of multiple watts.
It has been shown that state-of-the-art convolutional neural networks (ConvNets) consume over $1$W of processing power to achieve a desirable performance of $30$ frames per second (fps) with low-resolution QVGA frames on ASICs~\cite{azarkhish2018neurostream, pena2017benchmarking}.
The power consumption increases significantly with higher resolution inputs and higher frame rates. The power consumption reaches multiple watts on smartphones and can easily exceed $10$W with FPGA or GPU acceleration~\cite{cavigelli2015accelerating}.
To enable more exciting machine learning use cases, imaging systems need order-of-magnitude energy efficiency improvements to be able to analyze higher resolution image inputs while achieving higher frame rates.

This need for energy-efficiency has motivated recent investigations into 3-dimensional {\it stacked} integrated circuit architectures (Fig.~\ref{fig:near_sens_vision_pipeline}) to enable \emph{near sensor processing}.
In 2012, the first prototype of stacked sensors became available~\cite{hist3Dstacked}, enabling rudimentary image processing, such as demosiacing.
Other architectural trends propose processing near sensors; RedEye~\cite{likamwa2016redeye} and ShiDianNao~\cite{du2015shidiannao} perform ConvNet inference near the sensor, substantially improving energy efficiency of vision systems.
In combination of the two trends, future 3D-stacked sensors can layer the sensor, vision processor unit (VPU), and memory in the same package.

Unfortunately, sensor temperature sensitivity prevents a full adoption of near-sensor processing, creating noise in captured images.
In addition to generating less aesthetically pleasing images for human viewing, these noisy images reduce the visual task accuracy of computer vision applications~\cite{dodge2016understanding}.
Furthermore, poor lighting environments force the sensor to operate at high exposure and ISO\footnote{ISO controls the sensitivity of an image sensor to light.} for better scene capturing, which increases a sensor's vulnerability to noise. Despite a plethora of CPU dynamic thermal management (DTM) mechanisms~\cite{skadron2004temperature,kumar2008system,donald2006techniques,isci2006analysis,gomaa2004heat}, existing DTM techniques do not account for imaging requirements, turning a blind eye to the transient imaging needs of near-sensor processing.
Thus, despite performance and energy benefits of near-sensor processing, the temperature profile of visual computing limits stacked architectures in many situations.

While others have reported the thermal and noise implications of stacked-sensor processing~\cite{amir20183}, there lacks a comprehensive end-to-end modeling framework to allow  characterization studies considering energy, thermal, and noise implications of near-sensor processing workloads altogether. 
Thus, in \S\ref{sec:char}, we design and validate one such framework, aimed to assess thermal behavior of stacked-sensor architectures.
We validate our thermal models against real image sensor measurements and find that our RC models closely match with real measurements with an error margin of 0.1\%.
In addition to confirming and characterizing relationships between near-sensor processing power and sensor temperature, our modeling reveals a consequential insight: despite the coarse thermal time constant for the sensor to settle to steady-state temperatures, \emph{removing near-sensor power results in an immediate and significant reduction in transient junction temperature of the sensor}. For example, for a $2.5$ W system, the sensor temperature drops by roughly \SI{13}{\celsius} within $20$ ms when the processing is deactivated.

In \S\ref{sec:policies}, we build on characterized challenges to provision for imaging-specific temperature management.
We design the \textit{Stagioni} runtime to orchestrate temperature management for near-sensor processing.
We design, implement, and evaluate two temperature-aware scheduling policies as a part of the Linux-based Stagioni runtime -- {\it stop-capture-go} and {\it seasonal migration} -- for effective near-sensor vision processing. The policies aim to minimize system energy consumption and afford high-performance computation and high fidelity capture.
Stop-capture-go briefly suspends processing to allow for on-demand high fidelity captures and resumes the processing after the capture. On the other hand, seasonal migration occasionally shifts processing to a thermally isolated far-sensor processing unit for high fidelity capture.

In \S\ref{sec:eval}, we then use an emulation framework to evaluate the effectiveness of Stagioni's mechanisms to manage sensor temperature to suit imaging needs.
We make the following contributions:

\begin{itemize}
    \item {We develop and validate an end-to-end modeling framework for studying energy, thermal, and noise implications of near-sensor processing. We use this framework to characterize different implications of a typical 3D stacked near-sensor architecture.}
    \item{Motivated by the characterization findings, we design principles and propose novel fidelity-driven runtime mechanisms for effective sensor thermal management.}
    \item{Through our emulation-based evaluation, we show that for VPU power profiles that cause thermal problems, Stagioni determines the optimal amount of near-sensor task processing to avoid fidelity issues. By doing so, we find that Stagioni saves the average system power by 22-53\%; the actual savings depend on the power profile and image fidelity needs of the application.}
\end{itemize}

\noindent\textbf{Vision case study - Continuous life-logger: }
Enabling high performance and high efficiency near-sensor processing would unlock the potential for several vision/imaging applications, including sophisticated dashboard cameras, continuous augmented reality tracking, and other futuristic use cases.
Throughout this work, we study the implications of near-sensor processing and  evaluate the policies around a life-logger case study.
A wearable life-logger device chronicles important events and objects in a person's life.
The device runs object detection and tracking algorithms to continuously identify and locate objects in the scene.
Meanwhile, the device performs occasional captures upon detecting any important event, e.g., a person entering the scene.
This can form the basis for personalized real-world search engines, and assist those with memory impairments or visual impairments.

\section{Background and Related work}\label{bg}\label{bg_nearSensor}
\noindent\textbf{Near-sensor processing paradigm:}~~
The paradigm of near-sensor processing emerged in the 1990s to reduce the communication and storage overhead of off-sensor processing.
Early works~\cite{forchheimer1994near} leveraged physical properties to perform low-level image processing tasks, e.g., median filtering.
Later, researchers integrated image processing units~\cite{shi2005smart} after the read-out circuits in the imaging plane, outputting extracted image features. 
With advancement in 3D circuit integration,
recent works~\cite{amir20183} design 3D stacked image sensors, some of which include a system on a chip (SoC).
Inside the SoC, sensor, processor, and memory are stacked into the same package.
This architecture performs high-level image processing tasks, such as ConvNet-based classification.

3D stacked architectures have seen commercial advances.
For slow-motion capture, Sony~\cite{haruta20174} stack a DRAM beneath the sensor layers.
With local memory, the sensor captures and buffers frames at 1000 fps, sending them across the slower camera interface to the host.
Samsung~\cite{samsungstacked} uses a similar sensor for their recent Galaxy phone.
For surveillance, Sony~\cite{kumagai20181} integrated a motion estimation block, microcontroller, and DRAM in the 3D stacked sensor.

\vspace{0.5em}
\noindent\textbf{VPU architectures and power profiles:}~~
Though vision can be done through handcrafted feature analysis~\cite{lowe1999object}, the current trend uses ConvNets for visual tasks on a wide range of architectures.
High programmability, performance, and energy-efficiency are desired to meet the rapid pace at which ConvNets are evolving.

General-purpose platforms built around GPUs provide programmable high performance software libraries~\cite{jia2014caffe, abadi2016tensorflow} to implement ConvNets at the expense of more power, e.g., 60 fps at 10s to 100s of watts~\cite{nvidiagpus, pham2012neuflow, cavigelli2015accelerating}.
FPGAs provide performance and scalability at reduced power.
The state-of-the-art FPGA implementations~\cite{zhang2015optimizing, pham2012neuflow, zhang2016caffeine} typically consume several watts of power.
In recent years, we see the rise of domain specific processors such as Myriad2~\cite{pena2017benchmarking} that provide programmable SIMD capabilities on a RISC processor.
This brings down the power to a few watts~\cite{pena2017benchmarking}, but at the cost of performance, e.g., 3 fps.

Meanwhile, academic ASICs~\cite{chen2016eyeriss, han2016eie, du2015shidiannao} provide energy-efficiency and performance for ConvNets.
However, benefits are bottlenecked by DRAM accesses.
For example, Eyeriss~\cite{chen2016eyeriss} achieves 278 mW @ 35 fps for AlexNet.
But when scaled for VGG16~\cite{simonyan2014very}, performance drops to about 10 fps within the same power budget.

For reasonable performance, scalability, and mobility, the system power profile ranges from 1 to 15 W.
Placing these VPUs near the sensor and solving temperature challenges would unlock substantial improvements in performance and energy-efficiency through near-sensor processing.

\vspace{0.5em}
\noindent\textbf{Thermal noise in image sensors:}\label{bg_noise}~~
Image sensors are susceptible to different types of noise due to imperfections in lighting, sensing elements, and imaging circuitry.
Sources of noise can be grouped into fixed-pattern noise and temporal noise.
Fixed-pattern noise arises due to non-uniform sensitivities of photodiodes to light.
As it remains constant over time, conventional strategies read it once and subtract it later to eliminate its effect.
In contrast to fixed-pattern noise, temporal noise sources vary with each capture.

Temporal noise sources include read noise and dark current shot noise, which exhibit strong dependence on temperature.
All electronic noise sources, e.g., readout elements, amplifiers, are grouped together as read noise, which has a variance of kT/C.
This noise is due to random thermal activity of electronic charge carriers.
Dark current shot noise also stems from similar phenomenon happening in photodiodes; high temperatures trigger randomness in the photodiode charge carriers, thereby inducing more noise in images.
Unfortunately, thermal noise cannot be fully corrected using signal processing techniques without generating imaging artifacts \cite{marclevoy}.
The only solution is to manage sensor temperature.

\vspace{0.5em}
\noindent\textbf{Dynamic thermal management in microprocessors:}~~
For efficient thermal management, different techniques have been explored for multi-core processors. Stop-and-go~\cite{brooks2001dynamic} suspends the execution of a thread, for a while, when a core on which it is running gets overheated and resumes its operation once the core cools down. Heat-and-run~\cite{gomaa2004heat} technique migrates the thread from a  hotter core to a cooler one to allow the hotter core to cool down.
Traditional DTM techniques are designed to keep the processor power within a thermal design power (TDP).
We are inspired by the same core mechanisms -- stop-go and seasonal migration -- for power and temperature reduction.
In contrast to the existing works, we redesign these mechanisms to fulfill dynamic imaging needs.

\vspace{0.5em}
\noindent\textbf{Thermal problems in 3D stacked image sensor:}~~
Recent works report temperature issues in 3D stacked image sensors.
Amir et al.~\cite{amir20183} stack a DRAM and a deep neural network (DNN) processor beneath the sensor layer.
They report that sensor temperature can increase due to DNN computation, resulting in higher noise and lower ConvNet accuracy.
Lie et al.~\cite{lie2014analysis} report similar issues for their 3-layer stack architecture with a image compression unit integrated inside the stack.
Similar to earlier works, we report similar issues for our characterized 3D stacked image sensor.
However, previous works provide \emph{design time} solutions, e.g., statically partitioning computation to execute partial ConvNets on the sensor and the rest on the host.
Our work is complementary to theirs by providing \emph{runtime solutions} for thermal management.

\section{Modeling energy, thermal, and fidelity implications of near-sensor processing}\label{sec:char}
In this section, we construct a modeling framework to examine the implications of using stacked integration to place a VPU layered underneath the sensor for near-sensor processing.
Our estimates are based on a suite of parameterizable energy, temperature, and noise models of different hardware structures of a 3D stacked system.
To develop our models, we leverage datasheets, ITRS roadmaps, and commercial simulation software to produce accurate estimation of different thermal characteristics of 3D stacked sensors. We also validate these models through sensor hardware measurements.

{\bf Overview:} To better appreciate the insights offered by near-sensor processing, we study various system implications around our life-logger case study.
Our studies confirm that near-sensor processing minimizes off-chip data movement, thereby substantially reducing interface power and overall system energy consumption.
With near-sensor processing in our case study, we can reduce the system power of ResNet-based image classification by 52\%.
We also relate near-sensor processing power to image fidelity through temperature simulation, confirming that image fidelity degrades over time with additional near-sensor processing power.
We observe that removal of near-sensor processing power via throttling or computation offloading leads to rapid drops in sensor temperature, e.g., reducing temperature by \SI{13}{\celsius} in \SI{20}{\ms}.
We can exploit this observation to allow the sensor to operate at higher temperatures and lower image fidelities for energy-efficient vision, e.g., continuous object detection, while switching to low temperature operation for high-fidelity image capture when an application needs high quality photographs of a particular object.

\subsection{Energy analysis of near-sensor processing}
Near-sensor processing reduces energy-expensive data movement across the interconnects between different chips.
Here we examine energy profiles of vision pipelines, comparing traditional and near-sensor  pipelines.
Our energy models provide coarse estimation; actual numbers will depend on factors such as architectural decisions and patterns of execution.

\subsubsection{Energy of vision pipeline components}
Traditional pipelines operate across chips to connect a variety of subsystems: camera, processing unit, memory.
The camera chip connects to processing units on the System-on-Chip (SoC) through a standard camera serial interface (CSI) for data transfer and an I$^2$C interface for control and configuration.
Meanwhile, the SoC buffers image frames with DRAM through an external DDR interface.

\begin{table}[t]
\small
\caption {Energy-per-pixel of various components.
}
\label{tab:energy_per_component}
\begin{tabular}{lllll}
\hline
\textbf{Component}           & \textbf{Energy (pJ/pixel)} &  &  &  \\ \hline
Sensing                      & 595                        &  &  &  \\ \hline
Communication (Sensor - SoC) & 900                        &  &  &  \\ \hline
Communication (SoC - DRAM)   & 2800                       &  &  &  \\ \hline
Storage (Read)               & 283                        &  &  &  \\ \hline
Storage (Write)              & 394                        &  &  &  \\ \hline
\end{tabular}
\vspace{-3ex}
\end{table}

Using regression models on measurements and reported values, we construct a coarse energy profile model to motivate the need for near-sensor processing.
As shown in Table~\ref{tab:energy_per_component}, we find that sensing, processing, and storage consume 100s of pJ per pixel.
On the other hand, communication interfaces consume more than 3 nJ per pixel.

\emph{Sensing} requires an energy of 595 pJ/pixel~\cite{likamwa2013energy, choi2015energy}, mostly drawn from three components: pixel array, read-out circuits, and analog signal chain, which consume 25 pJ/pixel, 43 pJ/pixel, and 527 pJ/pixel, respectively.
\emph{DRAM storage} on standard mobile-class memory chips (8 Gb, 32-bit LPDDR4) draws 677 pJ/pixel for writing and reading a pixel value~\cite{micronspreadsheet}. This roughly divides into 283 pJ/pixel for reading and 394 pJ/pixel for writing.
\emph{Communication} over CSI and DDR interfaces incur 3.7 nJ/pixel, mostly due to operational amplifiers on transmitters and receivers.
We measure the interface power dissipation~\cite{xilinxpowercalc} on 4-lane CSI interfaces and LPDDR4 interfaces by inputting several data rates.
From this information, we construct a linear-regression model to estimate the energy per pixel to be 0.9 nJ/pixel over CSI and 2.8 nJ/pixel over DDR.
For \textit{computation}, we gather reported power dissipations of various ConvNet architectures from the literature.

\begin{figure}[t]
    \centering
    \begin{subfigure}[b]{0.46\columnwidth}
     \centering
        \includegraphics[width=.9\textwidth]{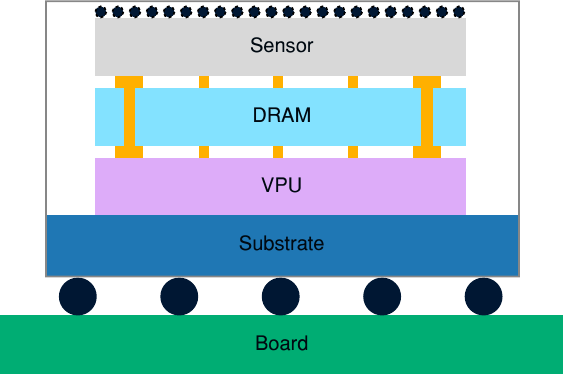}
        \caption{Cross-section of 3D stack.}
    \end{subfigure}
    \begin{subfigure}[b]{0.46\columnwidth}
    \centering
         \includegraphics[width=0.4\textwidth]{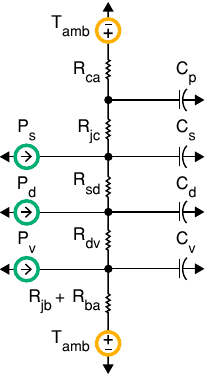}
         \caption{RC model of stack.}
    \end{subfigure}
\caption{Using the well-known duality between thermal and electrical phenomena, thermal modeling of stacked sensors can be performed by analyzing an equivalent RC circuit.}
\label{fig:stack_rc}
\end{figure}

\subsubsection{Energy of "far-sensor" processing architecture}
To illustrate the efficacy of near-sensor processing, in Table~\ref{tab:dnn_models}, we use our energy models to estimate the system power numbers of  traditional and stacked-sensor processing for different state-of-the-art ConvNet models.
We combine reported computation values with modeled sensing, storage, and communication energy to estimate the overall system power dissipation.
When operating at 1920 x 1080 at 34 fps, and using ResNet for inference on the SoC VPU, the modeled system consumes 2.7 W.

\subsubsection{Energy of near-sensor processing architecture}
On-chip data movement is known to be significantly more efficient than off-chip data movement by six orders of magnitude~\cite{borkar1999design}.
Advances in near-sensor processing leverage this for energy-efficiency gains, as shown in Fig.~\ref{fig:near_sens_vision_pipeline}.
Near-sensor processing moves the DRAM into the sensor to eliminate off-chip DDR movement, and moves the VPU into the sensor to reduce the CSI interface data rate.
Thus, the output of the sensor can be reduced from a few MB to a few bytes.
This information can be sent across efficient low data rate interfaces, e.g., I$^2$C.
Altogether, when applying our energy profile models to the near sensor processing pipeline, we find that the VPU near sensor system consumes 1.3 W, thereby yielding 52\% savings over traditional architectures.

\begin{table}[t]
\small
\caption {Thermal resistance and capacitance values of different components in RC model of stack.}
\label{tab:thermal_res}
\begin{center}
\begin{tabulary}{0.5\textwidth}{LLLLl}
\cline{1-2} \cline{4-5}
\textbf{Component} & \textbf{R (K/W)} &  & \textbf{Layer} & \textbf{C (J/K)} \\ \cline{1-2} \cline{4-5}
R$_{ca}$: Case-to-Ambient & 56 &  &  C$_{p}$: Package  & 1    \\
R$_{jc}$: Junction-to-Case & 6 &  &  C$_{s}$: Sensor   & 0.65m \\
R$_{sd}$: Sensor-to-DRAM & 0.6 &  &  C$_{d}$: DRAM     & 0.65m \\
R$_{dv}$: DRAM-to-VPU &    0.6 &  &  C$_{v}$: VPU     & 0.65m \\
R$_{jb}$: Junction-to-Board  & 40 &  &  &  \\
R$_{ba}$: Board-to-Ambient & 14 &  &  &
\end{tabulary}
\end{center}
\vspace{-1.5em}
\end{table}

\subsection{Thermal analysis of sensor processing}
Though tight integration yields energy efficiency and performance benefits, near-sensor processing generates heat at the sensor through thermal coupling between tightly integrated components.
While dynamic thermal management for CPU is only concerned with keeping peak power draw below a TDP, we pay close attention to temperature patterns, as transient temperature behavior affects image fidelity.
Conduction is the dominant heat transfer mechanism in integrated circuits.
To model temperature dynamics, we use simple thermal resistance-capacitance (RC) modeling~\cite{skadron2002control} techniques to determine stacked sensor characteristics.
\subsubsection{Deriving the component values in the RC model}
Fig.~\ref{fig:stack_rc} shows a typical structure of a 3D stacked sensor package and its RC model.
The sensor, DRAM, and VPU layers are stacked on top of each others, connected to each other, e.g., using through-silicon-vias.
The top of the stack opens to the surroundings through microlenses, while the bottom sits on a substrate that opens to the printed circuit board.
Mobile-class image sensors omit heat sinks or cooling fans, due to their size, weight, and placement challenges.
The layers consume power when active, which dissipates as heat.
We primarily consider vertical heat transfer; vertical resistances are several orders of magnitude smaller than the lateral resistances of convective heat transfer.
We obtain component values of the layers through analytical and empirical approaches.

Table~\ref{tab:thermal_res} shows different RC component values derived for our model.
Previous works report layer dimension values of typical 3D stacked image sensors~\cite{amir20183}.
In these works, the layer thickness ranges in the order of 1s to 10s of microns, while the layer's area ranges from 10s to 100s of mm$^2$.
The ITRS roadmap provides layer dimensions and material property constants $\rho$ and $c$ to define the guidelines for semiconductor fabrication.
From these, we derive the thermal resistance $R = \rho t/A$ and thermal capacitance as $C  = ctA$ where A is the layer's cross sectional area and t the thickness.

Package capacitance can be deduced empirically by observing the temperature trace of an image sensor chip while subjecting the sensor to thermal stress.
We construct regression models from the temperature trace of an OnSemi AR0330 smartphone-class image sensor to derive package capacitance.
Finally, termination thermal resistance depends on the properties of the casing and board.
Sensor companies make these values available through datasheets.
We use such provided values for typical packages directly in our model.\\

\noindent\textbf{Observation 1: Off-sensor power does not affect sensor temperature.}
While processing far from the sensor, the off-sensor VPU and SoC components do not influence the sensor temperature.
Even in tightly integrated systems, e.g., smartphones, the sensor and SoC reside on two different boards and communicate over a ribbon cable.
As a result, the sensor and SoC are nearly in thermal isolation.
That is, any increase in temperature of one component will not cause appreciable change in temperature of the other.
We verify this effect by running a CPU-bound workload on SoC on a Google Nexus smartphone while keeping the camera idle.
Our thermal camera instruments do not report any associated rise in camera temperature with an induced rise in SoC temperature.
Thus, in our study, we do not consider off-sensor thermal coupling effects.\\

\begin{figure}
	\centering
     \includegraphics[width=.85\columnwidth]{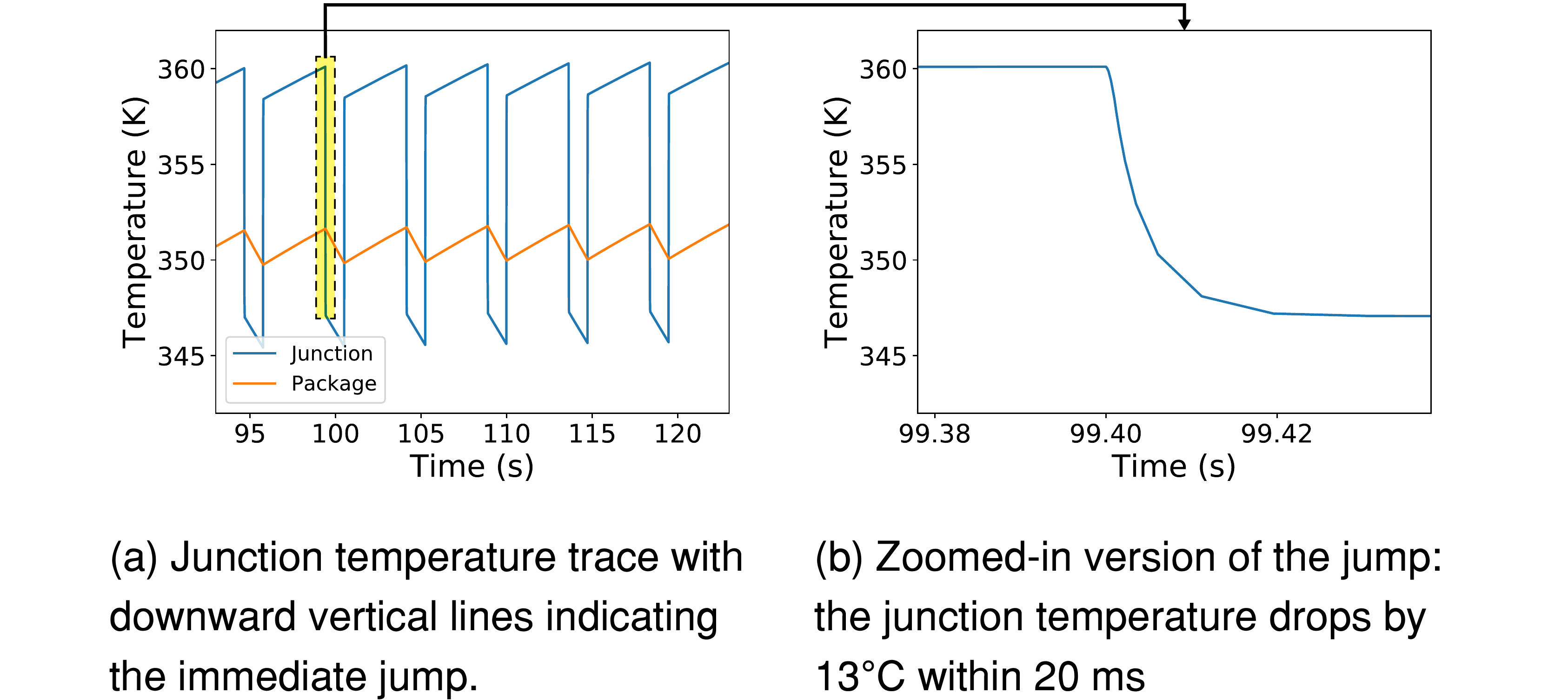}
     \caption{When disabling NSP, a rapid jump in junction temperature occurs within $20$ ms, due to junction time constants.}
     \label{fig:temperature_trace}
        \vspace{-1.5em}
\end{figure}

\subsubsection{Simulation-based thermal analysis}
Through LTSpice simulation on our RC models, we estimate the thermal behavior of near-sensor processing architectures.
We evaluate temperature profiles as the sensor operates in two different modes: \emph{NSP} mode, in which power dissipation is representative of capturing image frames and processing vision workloads near the sensor, and \emph{CAP} mode, in which power dissipations are representative of capturing image frames and either dropping frames or transmitting them to the SoC.
With various execution patterns, we can simulate the thermal behavior of the sensor as the system operates among different sensor modes.
Previous analysis has reported that we can safely ignore spatial variations in temperature if the chip power density is within \SI{20}{\watt\per\cm\squared}~\cite{itherm2018}, as is the case in NSP mode.
Power density, which is the power dissipated over the chip area, measures the degree of spatial non-uniformities in temperature.
The physical dimensions of our 3D stacked image sensor combined with the power profile of our case study results in a power density of \SI{16}{\watt\per\cm\squared}.
Therefore, we do not consider the spatial variations of temperature inside the stack for our modeling near-sensor processing architectures.

\textbf{Steady-state temperature: }
Inter-layer resistances are at least two orders of magnitude smaller than termination resistances.
This results in negligible drop across the resistor, leading to minuscule temperature gradients between layers.
For example, for \SI{1}{\watt} of VPU power, the sensor, DRAM, and VPU will be at \SI{60.7}{\celsius}, \SI{60.9}{\celsius}, and \SI{61.0}{\celsius}, respectively.
Thus, we combine the layers and model the sensor temperature as a single RC point.
Generally, reducing VPU power dissipation corresponds directly to temperature decrease.
The RC-based model shows that reducing near-sensor power from 1 W to 100 mW results in a temperature drop of \SI{5}{\celsius}.
Also, a higher ambient temperature leads to raised steady state temperatures.

\textbf{Transient temperature:} Thermal time constants govern the transient temperature of the stacked image sensor.
As the thermal capacitance of a chip package is often several orders of magnitude greater than that of a die, the thermal time constant of the package predominantly guides the trajectory of temperature to steady-state, taking 10s of seconds to reach a steady state temperature.
\noindent\textbf{Observation 2: The coarser thermal time constant allows dynamic temperature management policies ample time to form decisions, e.g., altering temperature by changing near-sensor power draw.}

Notably, near-sensor power dissipation raises the transient temperature of the sensor die above the package temperature.
This is because the heat source is on the sensor die itself, dissipating heat through the package into the ambient environment.
Consequently, reducing power dissipation rapidly reduces the gap between sensor die transient temperature and package temperature, as shown in Fig.~\ref{fig:temperature_trace}.
The speed of the temperature drop is governed by the sensor junction die time constant, which is on the order of milliseconds.
Prior works, e.g.,~\cite{skadron2004temperature,kumar2008system,donald2006techniques,isci2006analysis,gomaa2004heat}, have found similar temperature characteristics of the large, sudden drop for CPUs.
However, CPU thermal management can neglect fine-grained temperature variations  because its goal is to govern the processor junction temperature below a threshold.
Because transient temperature affects image fidelity, {\it these rapid temperature drops -- such as the charted \SI{13}{\celsius} drop in $20$ ms -- provide unique opportunities for dynamic thermal management for on-demand image fidelity}.
We further discuss this in \S\ref{sec:policies}.

\textbf{Validation: }
Due to the lack of configurable stacked sensors, we use an OnSemi's off-the-shelf mobile class camera~\cite{python1300} with an on-chip temperature sensor to validate our thermal insights.
We find that the validation results follow what our RC model predicts.
In particular, we collect the real-time temperature trace from the on-chip thermal sensor while configuring the sensor in preview and low-power modes.
We validate the relationship of temperature to power dissipation.
We observe that the sensor reaches a temperature of \SI{35.0}{\celsius} under a dynamic power dissipation of 250 mW.

When we use the power dissipation as input to our model, the steady-state temperature is estimated to be \SI{34.8}{\celsius}, within 0.06\% of real measurement. When we switch from the preview mode to the low-power mode, reducing the power dissipation to \SI{150}{\mW}, we observe a steady-state temperature of \SI{31.6}{\celsius}. The model prediction is \SI{31.4}{\celsius}, within 0.06\% of real measurement. 

We also observe the sudden temperature drop due to removal of near-sensor power dissipation. Upon transition to the lower power state, we see 30\% of the temperature reduction occurring within ~30 ms.
Our RC model predicts the junction time constant to be ~20 ms, which is close to what we observe through hardware measurement. We additionally validate the sudden temperature drop characteristic for higher power dissipation differences, leveraging a mobile SoC~\cite{sd636} and SnapDragon profiler~\cite{sd_profiler}, which also profiles battery power draw.
We notice substantial drop of \SI{15}{\celsius} in when there is a 3 W power removal by turning off a neural network based object detection application.

\subsection{Image fidelity implications of temperature}
While raised temperatures cause reliability and packaging issues for integrated circuits, they introduce another problem for image sensors: noise.
The influence of noise on vision tasks has been widely reported.
Dodge et al.~\cite{dodge2016understanding} find that neural networks have difficulty predicting semantics of an image when challenged by image noise.
Similar findings from Amir et al.~\cite{amir20183} find that image classification accuracy degrades with increase in temperature.
Thus, reliable vision demands images of reasonable fidelity.

Images for human consumption raise the fidelity bar for imaging; high fidelity is often needed in many real-life scenarios.
If a set of dashcam images is to be used in an auto insurance claim, the images need to have superior quality to obtain maximal information for decision-making.
While denoising can help mitigate fidelity issues, denoising algorithms often create imaging artifacts which also impair perceived image quality.
Thus, as images are required to fiducially represent the real physical world, imaging fidelity needs are more stringent than vision-based needs.

The sources of image noise are theoretically well-understood (\S\ref{bg}).
However, to understand the practical relationship between temperature and image quality on commercial sensors, we perform thermal characterization on a 3~Mp OnSemi AR0330 sensor~\cite{ar0330} connected to a Microsemi SmartFusion2 FPGA~\cite{imagingsolution}.
The AR0330 sensor includes noise correction stages inside the sensor, as is common in commercial sensors.
We use a heat gun to raise sensor temperature and capture raw images in a dark room setting while monitoring sensor temperature with a FLIR One thermal camera.

\begin{figure}
\centering
   \includegraphics[width=.7\columnwidth]{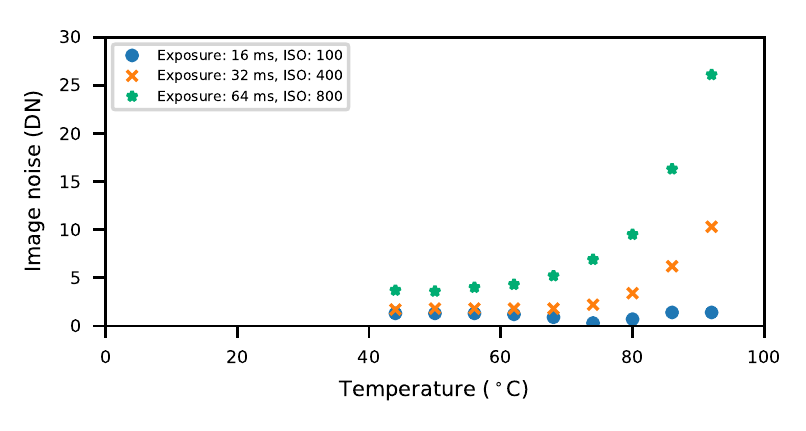}
   \caption{Variance of noise, expressed in pixel intensities, showing sensitivity to temperature, exposure, and ISO. }
   \label{fig:noise_vs_temp}
   \vspace{-1em}
\end{figure}

\begin{figure}
    \centering
    \begin{subfigure}[t]{0.3\columnwidth}
        \includegraphics[width=\textwidth]{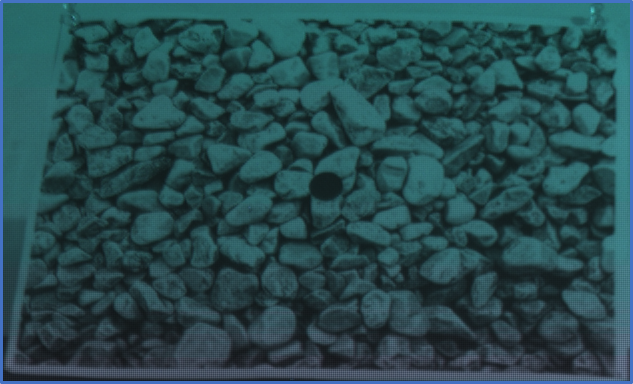}
        \caption{Image at \SI{44}{\celsius}}
        \label{fig:cooler_image}
    \end{subfigure}
    \quad
    \begin{subfigure}[t]{0.3\columnwidth}
        \includegraphics[width=\textwidth]{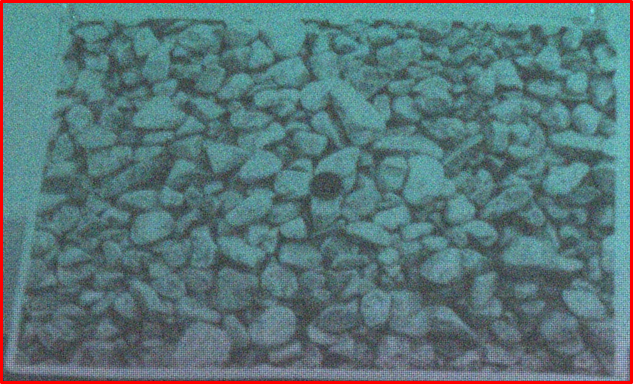}
        \caption{Image at \SI{92}{\celsius}}
        \label{fig:hotter_image}
    \end{subfigure}
    \quad
    \begin{subfigure}[t]{0.3\columnwidth}
        \includegraphics[width=.85\textwidth]{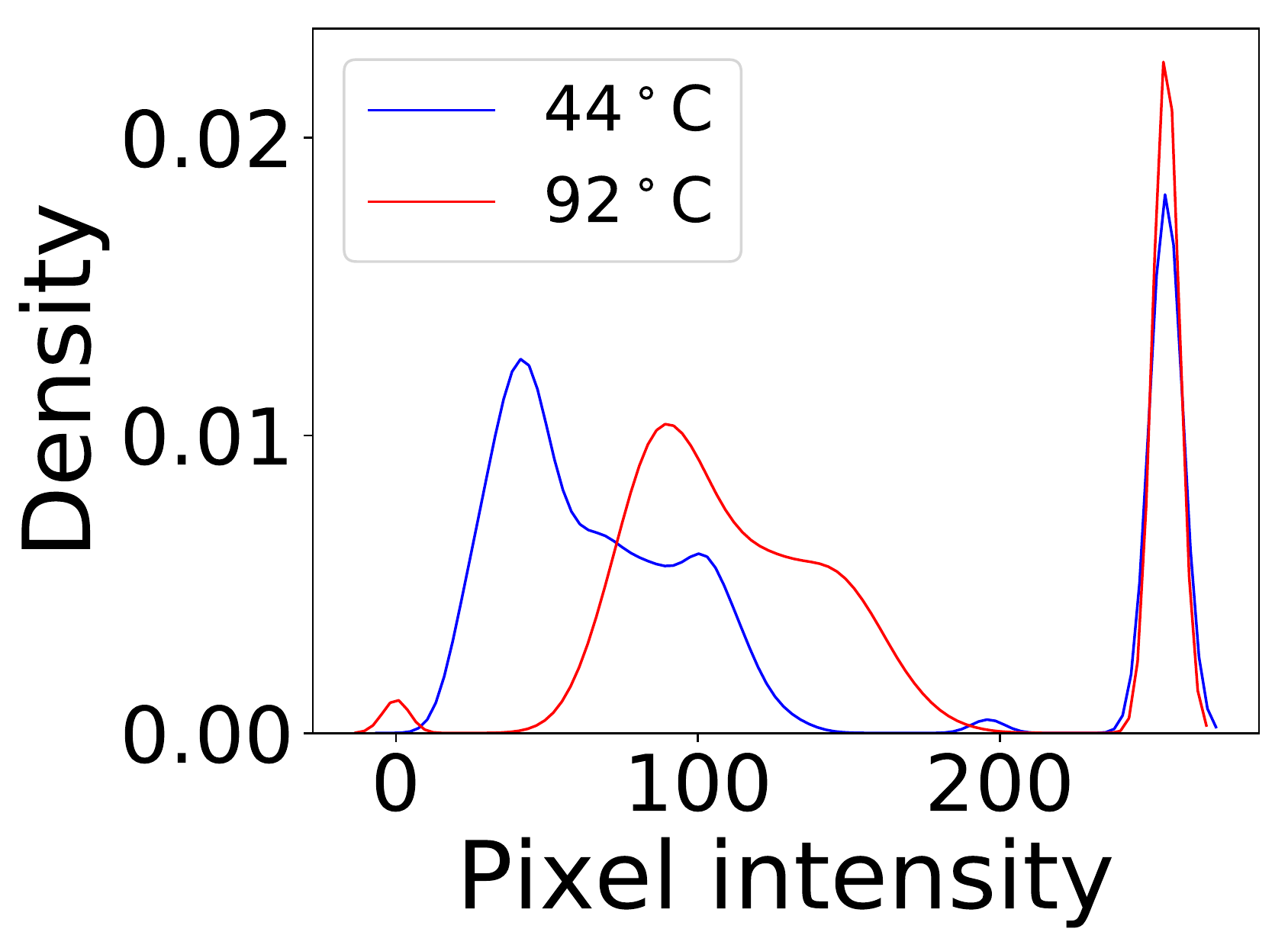}
        \caption{Image histograms}
        \label{fig:histograms}
    \end{subfigure}
    \caption{Two images captured at different temperatures and their histograms. The hotter image is brighter and grainier, due to the influence of thermal noise. This is also reflected in the shift in mean and variance width in the histogram. }
    \label{fig:image_n_histo}
\end{figure}

\subsubsection{Noise is more prevalent at high temperatures}
Fig.~\ref{fig:noise_vs_temp} charts a trend: sensors are particularly susceptible to noise above a particular temperature value.
This is despite the presence of noise correction stages inside the sensor.
We observe that the correction blocks bring the noise under control but only for lower temperature settings.
However, for high temperatures, the denoising fails to exercise control on noise minimization.
Notably, this knee shifts with exposure and analog gain settings, presumably due to noise amplification.
For instance, at high exposure and analog gain, which correspond to low light situations, sensors start to become thermally sensitive even at low temperatures, e.g., \SI{52}{\celsius}.
To adapt to experienced conditions, the sensor's thermal management should be adaptive to varying lighting conditions.

\subsubsection{Noise visibly and substantially impairs quality}
Thermal noise is visibly apparent on images, whether in low light or bright light conditions.
For example, Fig.~\ref{fig:image_n_histo} shows images captured under daylight conditions at different temperatures.
We can observe the graininess in the hotter image due to the strong influence of noise.
Paired with the noisy images, the histograms represent the pixel intensity distribution of an image.
The wider peaks in the distribution signify the variance of pixel intensity, while the mean of the peaks represent average intensity.
We can observe that the histogram of the hotter image shifts to the right, increasing pixel intensity due to dark current.
We also observe that the variance of the pixel intensity increases, due to increased thermal noise.

\subsection{Motivational observations}
To summarize, we have the following insights for NSP.
\begin{itemize}
  \item{Near-sensor processing architectures promote system energy-efficiency, but also increase sensor temperature}
  \item{Raised sensor temperatures aggravate thermal noise}
  \item{Smaller (ms) sensor junction time constants facilitate an immediate sensor temperature drop}
  \item{Fidelity needs are highly dynamic and depend on environment, e.g., lighting and ambient temperature}
  \item{Imaging demands more fidelity than vision}
\end{itemize}
These observations motivate the need for novel dynamic thermal management strategies for near-sensor processing.

\section{Thermal Management for Near-Sensor Processing}\label{sec:policies}
Our characterization shows that near-sensor processing increases system energy efficiency, but sacrifices image fidelity due to increased sensor temperatures.
This raises a natural question: \emph{Can we leverage near-sensor processing to create efficiency benefits while maintaining sufficient image fidelity for vision and imaging tasks?}
Driven by this, we develop novel mechanisms that can efficiently regulate sensor temperature for continuous and on-demand image fidelity needs.
In our design, these mechanisms are governed by a runtime controller, which we call \emph{Stagioni}.

Dynamic temperature management  for microprocessors is a mature research area, as we summarize in \S\ref{bg}.
However, traditional processor DTM mechanisms are not designed to suit imaging needs.
Rather than simply being limited by TDP, fidelity is impaired by the immediate transient sensor temperature while capturing.
Furthermore, thermal management for near-sensor processing should adapt to the situational needs of the vision/imaging application, e.g., allowing higher temperatures when in brighter environments and rapidly dropping temperature when high fidelity is required.

To account for near-sensor processing temperature management, we modify traditional DTM techniques to introduce two  mechanisms that quell image quality concerns, while striving to optimize for system power and performance.
(1) Stop-capture-go: Temporarily \textit{halt} near-sensor processing for thermal regulation and on-demand high fidelity capture.
(2) Seasonal migration: Occasionally \textit{migrate} processing to a thermally isolated far-sensor VPU for thermal regulation and on-demand high fidelity capture.

\begin{figure}
\centering
   \includegraphics[width=0.7\linewidth]{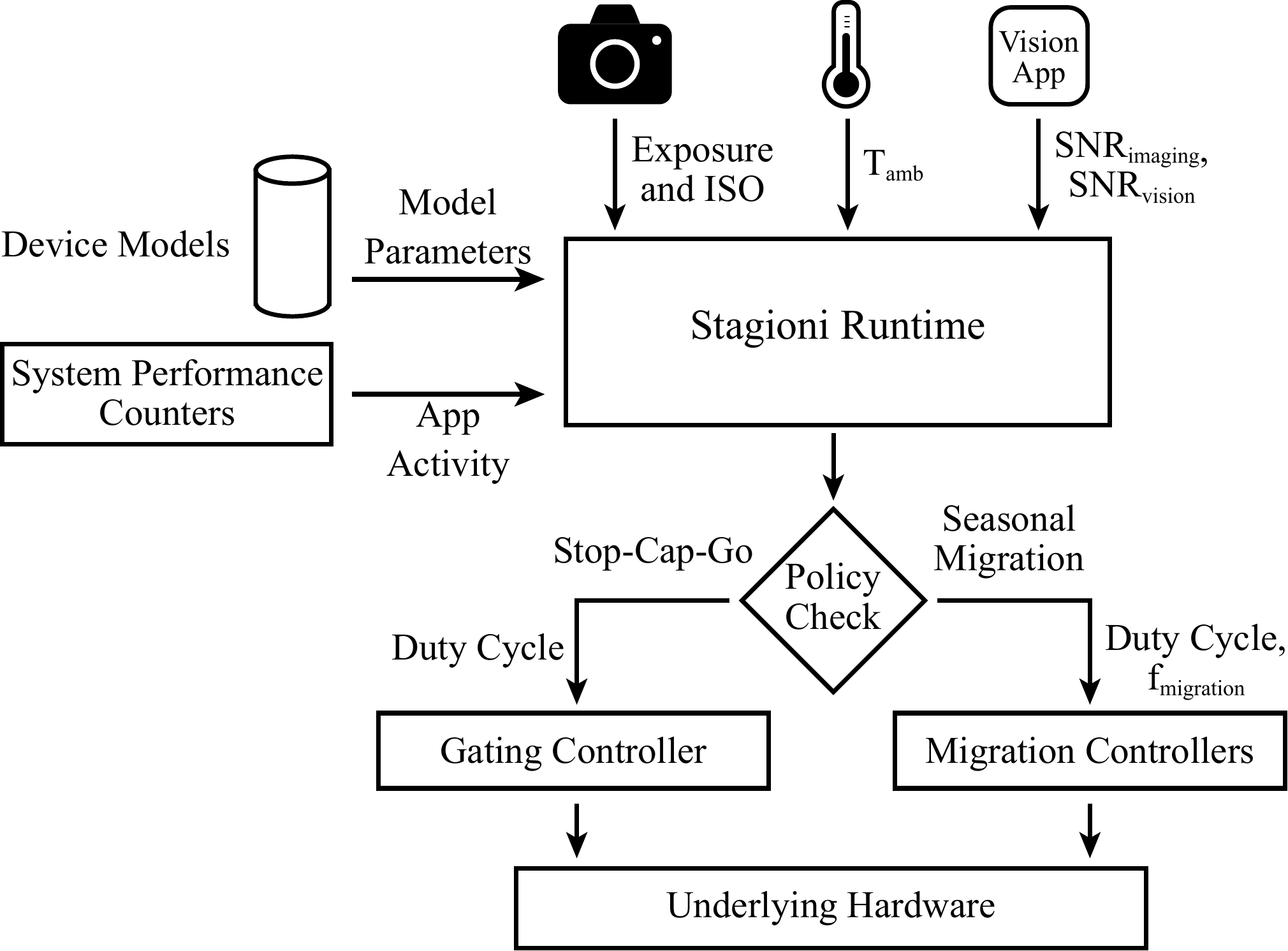}
   \caption{End-to-end execution flow of Stagioni.}
   \label{fig:flow_chart}
   \vspace{-1.5em}
\end{figure}

\subsection{Execution flow}
Here we describe the end-to-end execution flow (Fig. ~\ref{fig:flow_chart}) of Stagioni, starting from how it handles various inputs, operates on said inputs, and, eventually, generates the outputs which drive the policy controllers to enact different thermal management mecahnisms.
Stagioni primarily takes the fidelity needs and the ambient environment settings as inputs.
While the fidelity needs will be specified by the application developer through our API as we describe in the previous section, Stagioni leverages on-board sensors to derive the ambient settings.
For example, Stagoni uses  an ambient temperature sensor typically available on mobile phones to derive its value.
Along similar lines, it obtains the ambient lighting situation by reading the exposure and ISO values from automatic exposure controller module available in phone cameras.

In addition to fidelity and ambient parameters, Stagioni also leverages on-system performance counters to estimate the application activity information such as the number of memory loads and stores and number of arithmetic instructions.
This information is processed against the characterized models, stored in the system memory as look-up tables, to derive power, temperature, and noise trends.
These trends along with the fidelity and ambient constraints will constitute different thermal boundaries of the system.

Based on these thermal boundaries, Stagioni analytically determines different policy parameters such as duty cycle and migration frequency.
Finally, these policy parameters are fed to appropriate policy controllers -- gating controller for stop-capture-go and migration controller for seasonal migration -- to put the thermal management mechanisms into action.

\subsection{Design principles for sensor thermal management}
To design thermal management mechanisms that are effective for near-sensor processing, we introduce three core principles:
(1)~Situational temperature regulation: The mechanism should confine sensor temperature within a threshold that suffices for imaging fidelity needs.
(2)~On-demand fidelity: Upon application request, the mechanism should quickly drop the temperature to desired capture temperature for high fidelity imaging.
(3)~Duty cycle governs system efficiency.
Here, we discuss these in detail.

\subsubsection{Situational temperature regulation}
As we discuss in \S\ref{sec:char}, vision tasks have varying fidelity needs, which are sensitive to camera settings, e.g., ISO and exposure, and lighting situation, e.g., bright conditions.
This translates directly to a simple upper bound for temperature:
{\small
\begin{equation}
T_{sensor} < T_{vision}
\end{equation}
}
Thus, temperature management must be cognizant and respectful of immediate vision task requirements in situational conditions to provision for effective vision accuracy.

\subsubsection{On-demand fidelity}
While vision processing can operate on low fidelity images, certain applications may require high fidelity images on demand, e.g., life logging capture after object detection.
Such capture must be immediate, before the object leaves the view of the camera.
Fortunately, as we characterized, sensor temperature rapidly drops with the removal of near-sensor power, i.e., by entering $CAP$ mode.
For example, when the sensor drops its near-sensor power consumption from \SI{2.5}{\watt} to \SI{100}{\mW}, the sensor drops in temperature by \SI{13.2}{\celsius}.
We experimentally observe that sufficient temperature drop (98.2\%) can be achieved within a time of four time constants, which we define as $t_{jump} = 4\times RC_{die}$.
In our simulation, this amounts to 20 ms.
Temperature management can leverage this drop to provision for on-demand high fidelity.

The temperature drop is directly proportional to the disparity between the near-sensor power before and after power reduction: $T_{jump} = \alpha (P_{NSP} - P_{CAP})$.
We find that for our modeled sensor, every \SI{1}{\watt} causes a \SI{5.5}{\celsius} temperature jump, i.e., $\alpha = \SI{5.5}{\celsius\per\watt}$.
When constrained by a latency deadline, e.g., to immediately capture a moving object or to meet a synchronization deadline, the achievable jump within the latency deadline is a fraction of the time it takes to drop: $T_{jump}^{latency} = T_{jump}\times (e^{-t_{latency}/RC_{die}})$
Thus, to provision for predicted fidelity needs and latency needs of an application, the temperature management mechanism can set reduced bounds:
{\small
\begin{equation}
T_{sensor} < T_{imaging} + T_{jump}^{latency}
\end{equation}}
\vspace{-2em}
\subsubsection{System power minimization through duty cycle}
While removal of processing power can regulate temperature and provide on-demand high fidelity captures, the scheduling of operation should also strive to optimize for average system power. We can characterize this through the duty cycle and frequency of switches between \emph{NSP} and \emph{CAP} modes.
For duty cycle $d$, switching frequency $f_{switch}$ and energy per switch $E_{switch}$, average system power can be modeled as:
{\small
\begin{equation}
\label{eqn:syspower}
P_{avg} = d \times P_{NSP}^{system} + (1 - d) \times P_{CAP}^{system} + f_{switch} \times E_{switch}
\end{equation}}
In minimizing average power, there is a notable tradeoff between the duty cycle and the frequency of switches. Spending more time in $CAP$ mode allows the sensor to cool down, increasing the length of time spent in $NSP$ mode as well.
On the other hand, spending less time in $CAP$ mode allows the sensor to spend a greater proportion of time in $NSP$ mode, promoting energy savings through the duty cycle, at the expense of number of switches.
Notably, the time spent in each mode must be a multiple of time spent capturing an image. It is not possible to switch to $CAP$ mode for a partial frame duration while an image is being captured.
For our implementation, which has minimal switching overhead, higher duty-cycles tend to provide favorable average system power profiles.

\subsection{Stop-capture-go}
The traditional stop-go DTM technique regulates processor temperature by halting execution through clock gating.
For near-sensor processing, we can similarly put the sensor in \CAP mode, gating near-sensor units for some time before resuming \NSP mode.
The resulting "temporal slack" allows the sensor to regulate capture fidelity at the expense of task performance.
Stop-go is architecturally simple, requiring only the ability to gate the clock or power of components.

Unlike traditional stop-go, our proposed stop-capture-go requires unique modifications for near-sensor processing.
First, frequently clock gating the entire sensor is not advisable; interruptions to the camera pipeline create capture delays on the order of multiples of frames.
Instead, the system will clock gate the near-sensor VPU and DRAM, putting the sensor into CAP mode.
Second, rather than being governed by TDP, the temperature regulation will trigger as the sensor reaches a situational upper bound specified by the design principles, such that {\small $T_{sensor} <T_{vision}$} and {\small $T_{sensor} < T_{imaging} + T_{jump}^{latency}$}.
Third, the application can request an execution halt to achieve on-demand fidelity. For this, the sensor enters CAP mode to retrieve the frame.

\subsubsection{Parameterization of stop time}
The amount of "stop" time -- the amount of time the processor is halted -- is an important policy parameter.
During the stop time, the system will "drop" frames, failing to process them, although they may be captured.
Elongated stop times allow a sensor to cool down, reducing the number of switches.
Stop times can be detrimental, as contiguously dropped frames may contain important ephemeral visual information.
Thus, if a system wishes to prioritize a continuity of visual information, stop time should be reduced.
In our simulated study, we find that the minimal stop time of $33$ ms (one frame time) is sufficient to cool the sensor from \SI{87} to \SI{74}{\celsius}, enabling sufficient temperature regulation and on-demand fidelity.

\subsubsection{Usability of stop-capture-go}
Due to the architectural simplicity of stop-capture-go, system overhead is minimal, promoting continuously low system power.
However, frequent frame drops impair visual task performance.
Thus, stop-capture-go is suitable for systems that demand low power but are not performance-critical and/or systems that require minimal architecture modifications.

\subsection{Seasonal migration}
While stop-capture-go is a simple policy for temperature regulation and high-fidelity captures, it degrades application performance by halting execution.
Towards minimizing performance loss, we investigate seasonal migration for near-sensor processing.
Seasonal migration shifts the processing to a thermally isolated computational unit, allowing continuous computing.
As we model in \S\ref{sec:char}, spatial thermal isolation between the sensor and SoC allows thermal relief.
Enabling seasonal migration comes at the expense of duplicated computational units near and far from the sensor, but effectively regulates temperature without sacrificing performance.

\begin{figure}
\centering
   \includegraphics[width=.7\columnwidth]{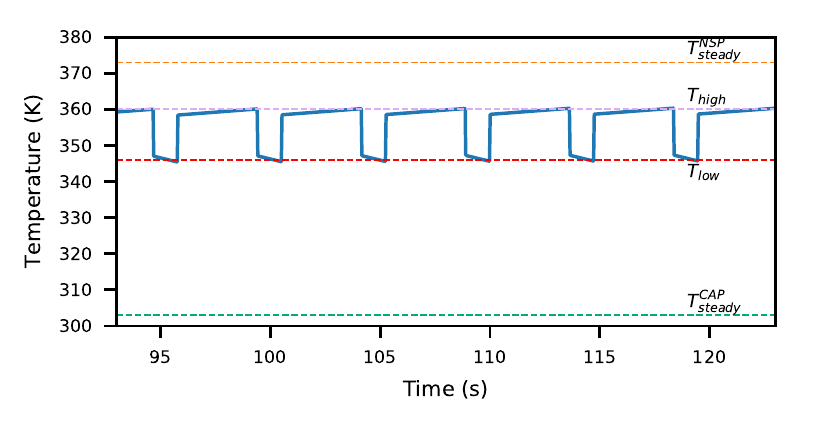}
   \caption{Transient response of seasonal migration mechanism with 77\% duty cycle to confine sensor temperature within thermal boundaries ($T_{high}$ and $T_{low}$).}
   \label{fig:temperatureTraceWithBounds}
   \vspace{-1.5em}
\end{figure}

As shown in Fig.~\ref{fig:temperatureTraceWithBounds}, seasonal migration is governed by two temperature boundaries: $T_{high}$ and $T_{low}$.
In efficiency phase, triggered when the sensor reaches a temperature below $T_{low}$, it will enter NSP mode, performing near-sensor processing for system efficiency.
In cooling phase, triggered when the sensor reaches a temperature above $T_{high}$, it will enter CAP mode, performing off-sensor processing on the SoC, allowing the sensor to cool down.
The alternation between phases allows the system to balance efficiency with temperature.
For on-demand fidelity, the system enters the cooling phase regardless of sensor temperature.

\subsubsection{Parameterization of thermal boundaries}
$T_{high}$ and $T_{low}$ are important policy parameters, controlling the balance of efficiency and temperature.
$T_{high}$ forces sensor temperature regulation, and thus should be set to shift to situational needs:
{
\small
\[T_{high} = min(T_{vision}, T_{imaging} + T_{jump}^{latency})\]
}
Meanwhile, the gap between $T_{high}$ and $T_{low}$ controls the system efficiency implications of the policy.
Because it takes more time for the sensor temperature to bridge a larger gap, larger gaps decrease the frequency of switches, while smaller gaps increase the frequency of switches.
The $T_{high} - T_{low}$ gap also controls the duty cycle of the system. When the desired sensor temperature range is closer to steady-state NSP temperature than steady-state CAP temperature, smaller gaps produce favorable duty cycles, spending more time in NSP mode.
As shown in Eqn.~\ref{eqn:syspower}, the average system power is a function of this duty cycle, balanced against the energy overhead and frequency of switches.
Thus, $T_{low}$ should be chosen to create a gap that optimizes average system power.

As we defined earlier, the duty cycle is the proportion of time spent in $NSP$ mode.
For seasonal migration, the relationships can be derived from standard charging models.
After the rapid drop or rise in temperature $T_{jump}$, which takes approximately $t_{jump}$ amount of time, the sensor follows an RC charging curve towards the steady state temperature of the $NSP$ or $CAP$ mode.
Altogether, this can be used to model duty cycle $d$ and frequency of migration $f_{migration}$.
{
\small
\[t_{warming} = RC \times \ln\bigg(\frac{T_{steady}^{NSP}- (T_{low} +T_{jump})}{T_{steady}^{NSP} - T_{high}}\bigg) + t_{jump}\]
\[t_{cooling} = RC \times \ln\bigg(\frac{(T_{high} - T_{jump}) -  T_{steady}^{CAP} }{T_{low} - T_{steady}^{CAP} - T_{jump}}\bigg)+ t_{jump}\]
\[d = t_{warming}/(t_{warming} + t_{cooling}) \]
\[f_{migration}= 2/(t_{warming} + t_{cooling})\]}
\vspace{-1.5em}

\subsubsection{Usability of seasonal migration}
Depending on implementation, seasonal migration could suffer from the switching latency and energy overhead resulting from state transfer and synchronization in shifting processing from one computational unit to another.
However, reducing this migration overhead is a well-studied problem in distributed systems~\cite{milojivcic2000process}.
Several reported techniques mitigate migration latency, e.g., pre-copy-based migration~\cite{richmond1997new}, which promote smooth execution performance while incurring energy overhead by keeping both computational units on while preparing for migration.
Similarly, in our implementation, prior to migration, we prepare the system by pre-emptively starting up the target computational unit and initiating its context so it is prepared for execution.
Consequentially, there is only a minimal switching overhead of \SI{100}{\us}, which is negligible in comparison to ms-scale image capture times.

\subsection{Stagioni Runtime Controller}
We propose the Stagioni Runtime Controller to execute the thermal management at runtime.
Stagioni's responsibility is to guarantee the fidelity demands of the application, coordinating state transfer between the operating modes to ensure smooth transition.
Stagioni could be designed in a multitude of ways, e.g., a dynamically linked library, a runtime OS service, or dedicated hardware.
In our implementation and evaluation, Stagioni is a runtime OS service that sits on the near-sensor processor, allowing the SoC to sleep.
Many migration controller designs would sufficiently and equivalently serve the purposes of decision-making.
Here we describe one set of modules that would achieve the goals.

\textbf{API for application-specific fidelity needs: }
A vision application only needs to provide three pieces of information to the controller:
(1) continuous image fidelity requirement for vision
(2) on-demand image fidelity requirement for imaging
(3) when to trigger on-demand fidelity.
A simple API can enable developers to specify requirements from their applications. A class with the following methods would suffice:
\begin{itemize}
  \item{\texttt{setVisionSNR(float)}: specify continuous fidelity}
  \item{\texttt{setImagingSNR(float)}: specify on-demand fidelity}
  \item{\texttt{triggerOnDemandFidelity()}: request high fidelity }
\end{itemize}

Stagioni translates expectations into thermal management, sidestepping any form of developer burden.
To do this, the controller applies application-specific requirements into appropriate policy parameters through characterized device models.
Stagioni also continuously adapts policy parameters to situational settings, i.e., ambient temperature and lighting, to meet ongoing constraints.

Stagioni orchestrates the execution pattern in runtime, which consists of several system-level events.
For stop-capture-go, Stagioni would use simple power gating mechanisms such as clock gating.
For seasonal migration, Stagioni would handle the communication between two chips.

To this end, Stagioni can use simple message passing schemes to synchronize states between the sensor and the host.
One such scheme, implemented in our evaluation, could operate as follows:
(\emph{i}) The temperature monitor detects a thermal trigger and raises an interrupt.
(\emph{ii}) Stagioni sends a signal to the SoC controller to prepare for migration.
(\emph{iii}) In return, the SoC controller starts the application and sends an acknowledgement to the source conveying that it is ready to accept the tasks.
(\emph{iv}) Stagioni then transfers application context data transfer from source's memory to the host's memory.
(\emph{v}) Once the data transfer is done, both migration handlers notify their corresponding applications.
The offloaded tasks run in the new context loading the state from the memory.
This sequence of steps can be scheduled prior to the migration event, such that immediate migration is possible.

\begin{table}[t]
\small
\center
\caption{DNN models and corresponding power profiles }
\begin{tabular}{|c|c|c|c|}
\hline
\textbf{\begin{tabular}[c]{@{}c@{}}DNN model\\ (VPU arch)\end{tabular}} & \textbf{\begin{tabular}[c]{@{}c@{}}Frame rate \\ (fps)\end{tabular}} & \textbf{\begin{tabular}[c]{@{}c@{}}Trad. Sys \\ Power (W)\end{tabular}} & \textbf{\begin{tabular}[c]{@{}c@{}}NSP Sys \\ Power (W)\end{tabular}} \\ \hline
\begin{tabular}[c]{@{}c@{}}AlexNet\\ (Myriad2)\end{tabular}             & 12                                                                   & 3                                                                       & 1.86                                                                  \\ \hline
\begin{tabular}[c]{@{}c@{}}mobileNetSSD\\ (Myriad2)\end{tabular}        & 11.8                                                                 & 1.92                                                                    & 0.9                                                                   \\ \hline
\begin{tabular}[c]{@{}c@{}}GoogLeNet\\ (Neurostream)\end{tabular}       & 83                                                                   & 3.13                                                                    & 1.81                                                                  \\ \hline
\begin{tabular}[c]{@{}c@{}}ResNet50\\ (Neurostream)\end{tabular}        & 34                                                                   & 2.67                                                                    & 1.34                                                                  \\ \hline
\end{tabular}
\label{tab:dnn_models}
   \vspace{-1.5em}
\end{table}

\section{Experimental Methodology}
Since there are no readily available off-the-shelf programmable 3D stacked image sensors, we use emulation techniques to implement Stagioni's mechanisms.
Our emulation framework operates on our characterized energy, noise, and thermal models and reports system metrics such as system power and performance.
We design and implement Stagioni as a runtime controller and integrate it into the emulation setup to study execution patterns of different policies.

\subsection{Emulated architecture}
We model a 3D stacked sensor architecture in our emulation framework.
For its sensing element, we emulate the fidelity characteristics of an AR0330~\cite{ar0330} which is a typical mobile-class image sensor with sufficient number of pixels for providing high-quality images.
For its storage element, we emulate the power profile of a 4 Gb LPDDR4 DRAM, which is commonly seen in commercial 3D stacked sensors~\cite{haruta20174} for slow-motion video capture.
Finally, for its processing element, we emulate the power characteristics of a Myriad2, a vision co-processor found in mobile devices~\cite{googleaiy,intelncs}, capable of neural network processing and feature-based processing, and also Neurostream~\cite{azarkhish2018neurostream}, another recent candidate architecture for energy-efficient vision processing.
In our emulation, the resulting stacked sensor connects to an ARM-based mobile-class SoC through a standard 5 Gbps CSI interface.
We assume that the SoC also contains a vision co-processor, i.e., Myriad2/Neurostream, to which it can offload the tasks.

\subsection{Emulation setup}
While we can use our modeling to evaluate thermal and energy behavior, the runtime behavior of Stagioni and its adaptiveness to different ambient conditions can only be assessed through hardware and software implementation.
While we could study the mechanisms on any mobile platform such as SnapDragon and TX2, we choose an FPGA platform because it has programmable fabric where we can synthesize a neural processing unit to emulate a vision co-processor.
To this end, we build a FPGA-based emulation platform based off two ZCU102 boards.
One of them emulates the stacked sensor, while the other emulates the SoC.
Both employ hardware-accelerated vision processing through the CHaiDNN library~\cite{chaidnn}.
We use 1 Gbps Ethernet for communication, simulating a standard CSI interface that has similar bandwidth characteristics.

The Stagioni controller takes the type of policy and associated model parameters as inputs.
The parameters generate a temperature-dependent mode schedule that governs task execution at runtime.
The controller also handles high fidelity requests and services them to deliver high quality images through appropriate mechanisms.
During $CAP$ mode for stop-capture-go, the controller gates the execution of the neural network invocation.
For seasonal migration, the controller performs message passing over Ethernet for state transfer and implements producer-consumer queues for synchronization.
During $NSP$ mode, the controller gates the SoC FPGA.


\begin{figure}[t]
\centering
   \includegraphics[width=.8\columnwidth]{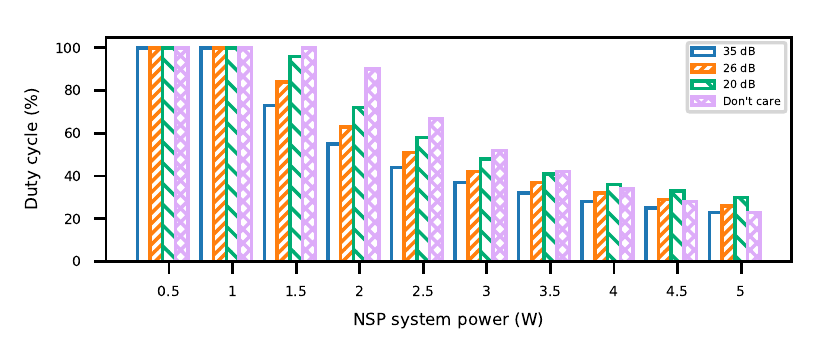}
   \caption{Influence of fidelity and NSP power on duty cycle.}
   \label{fig:dutyCycleVsappsVsSNR}
   \vspace{-1.5em}
\end{figure}

\subsection{Workloads}

\textbf{Vision tasks:}
We evaluate the life-logger use case which performs continuous vision with occasional imaging upon detection of interesting events.
For our vision tasks, we study two forms of vision: (i) image classification, identifying scenes, and (ii) object detection, locating objects in a scene.
For each of these tasks, as shown in Table~\ref{tab:dnn_models}, we choose a variety of the state-of-the-art DNN models with different input, memory, and computational requirements to stress different elements in our stacked architecture.

\textbf{Metrics and policies:}
The major objective for evaluating a policy is to regulate temperature for capture fidelity, while optimizing system power with minimal performance overhead.
We use SNR to gauge image quality and frame drops for performance overhead.
In addition to stop-capture-go and seasonal migration, we consider full-far sensor processing (status quo) for comparison.

\textbf{Fidelity choices:}
In imaging, a SNR of 20 dB~\cite{snr_imaging_wiki} is considered as acceptable quality under well-lit conditions.
However, the bar is higher for more challenging conditions, including environments where fiducial detail is important.
This can be seen in sensor data sheets~\cite{ar0330} where manufacturers design cameras to deliver higher SNR values, e.g., 35 dB  for excellent performance under low-light conditions.
Therefore, to capture all real-world scenarios, we use range of fidelity choices \{35 dB, 26 dB, 20 dB\}, and a "don't care" scenario in which the application continuously performs vision without any on-demand high fidelity imaging.

\textbf{Environment conditions:}
We evaluate a wide range of temperature and lighting conditions.
For evaluating ambient temperature effects, we use values from \SI{20}{\celsius} to \SI{40}{\celsius}.
Meanwhile, lighting translates into different camera settings, i.e., exposure and ISO.
We use the flexible CapraRawCamera~\cite{caprarawcamera} camera app to automatically determine appropriate camera settings based on the scene lighting.
We use the following camera settings for three sensor illuminations.

\begin{itemize}
  \item{Outdoor daylight (32000 lux): Exp.: $16$ ms, ISO: 100}
  \item{Indoor office light (320 lux): Exp.: 32 ms, ISO: 400}
  \item{Dimly lit office light (3.2 lux): Exp.: 64 ms, ISO: 800}
\end{itemize}
\vspace{-1.5em}

\section{Evaluation Results and Analysis}\label{sec:eval}

\noindent{\bf Result Summary:} We investigate the effectiveness of our proposed policies in meeting fidelity demands of various vision tasks around the life-logger use case.
For our evaluated tasks, we find that our policies deliver system power savings ranging from 22\% to 53\%.
The actual savings vary with the fidelity requirements and power profile of the workload.
The savings stem from maximizing near-sensor task operation, which helps reducing system power by cutting down energy-expensive off-chip data movements.

We also find that Stagioni improves system energy efficiency without much performance loss.
For seasonal migration, overhead due to offload can be contained within \SI{0.1}{ms} through practical techniques, e.g., pre-copy, thereby leading to no frame drops.
On the other hand, stop-capture-go suffers from occasional frame drops, leading to performance loss from the duty cycle of the system.

In addition, we study Stagioni's adaptiveness to different dynamic ambient conditions, such as lighting and ambient temperature.
We find that Stagioni quickly and smoothly adapts the thermal boundaries based on ambient conditions.

\begin{figure*}[t]
   \centering
   \begin{subfigure}[t]{0.2\linewidth}
    \includegraphics[width=\textwidth]{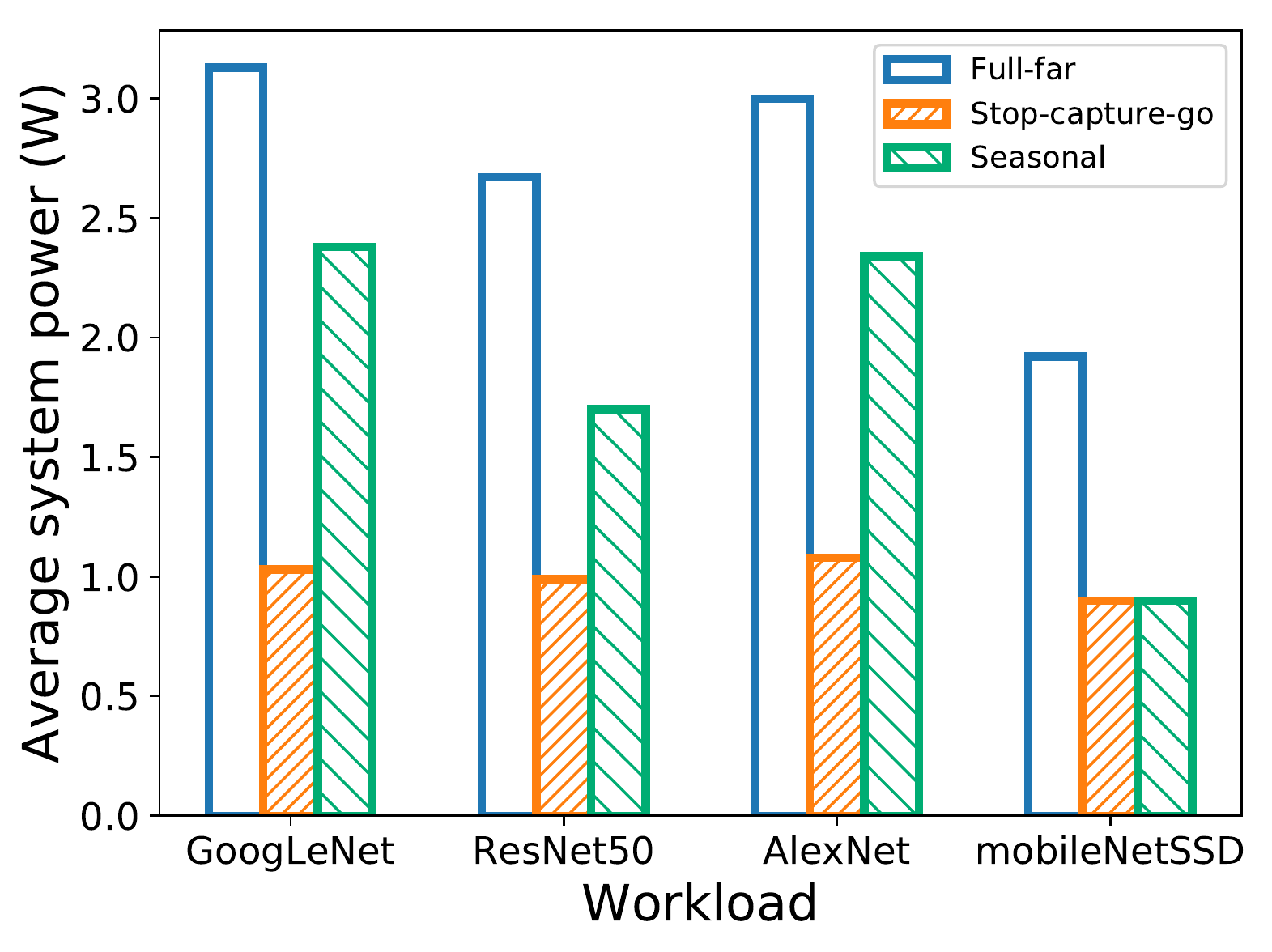}
    \caption{35 dB}
   \end{subfigure}
   \quad
   \begin{subfigure}[t]{0.2\linewidth}
    \includegraphics[width=\textwidth]{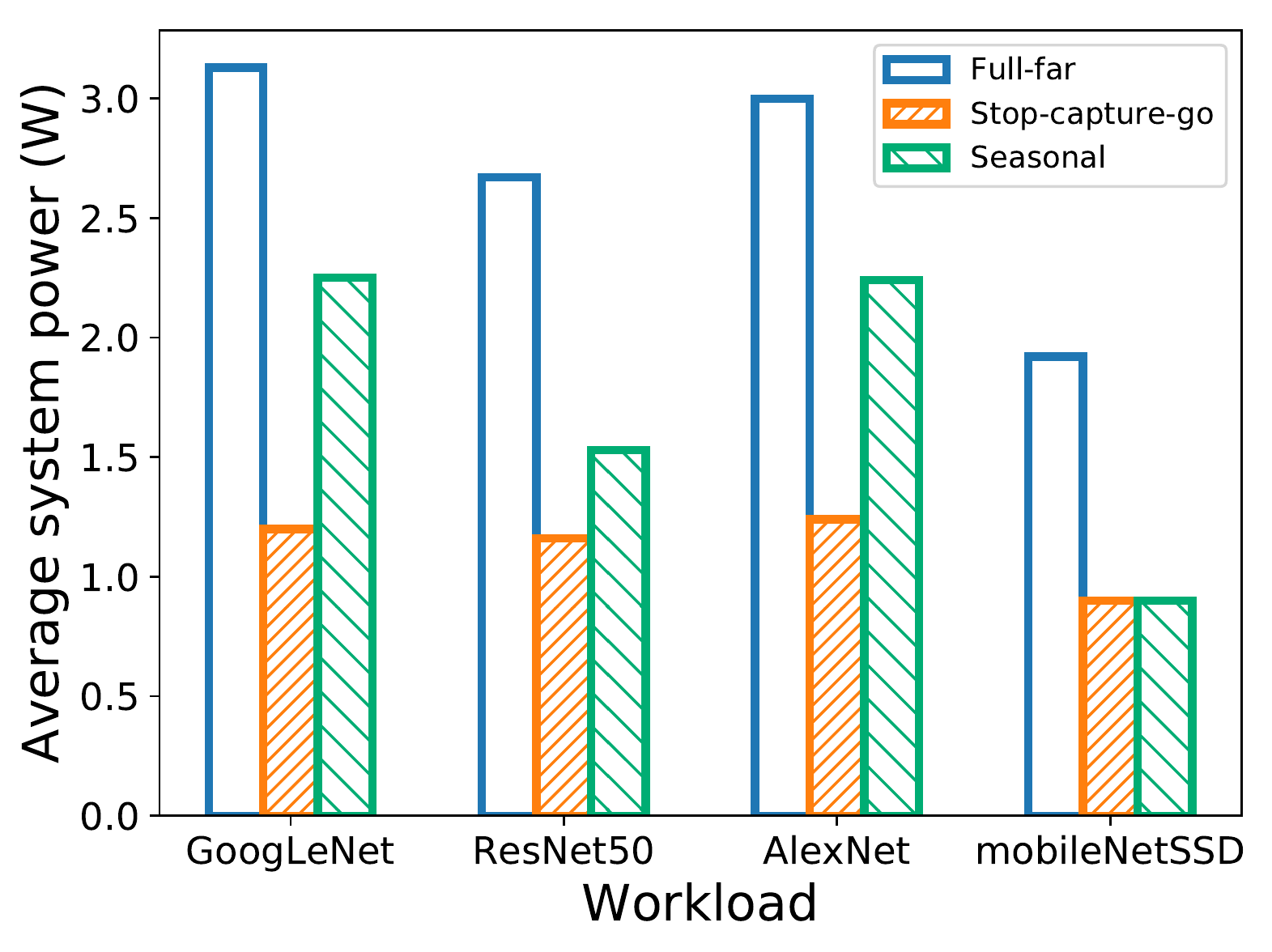}
    \caption{26 dB}
   \end{subfigure}
   \quad
   \begin{subfigure}[t]{0.2\linewidth}
    \includegraphics[width=\textwidth]{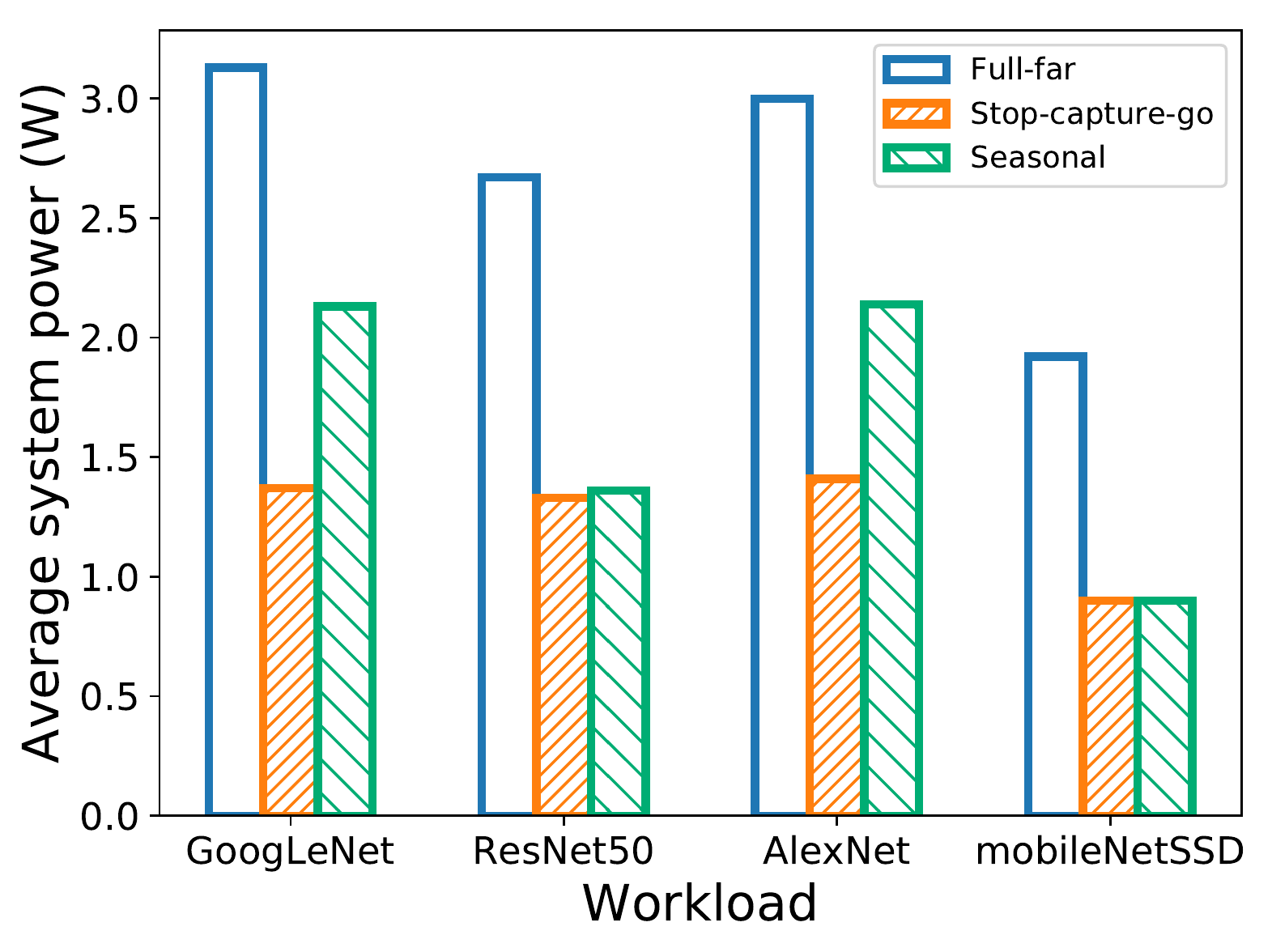}
    \caption{20 dB}
   \end{subfigure}
   \begin{subfigure}[t]{0.2\linewidth}
    \includegraphics[width=\textwidth]{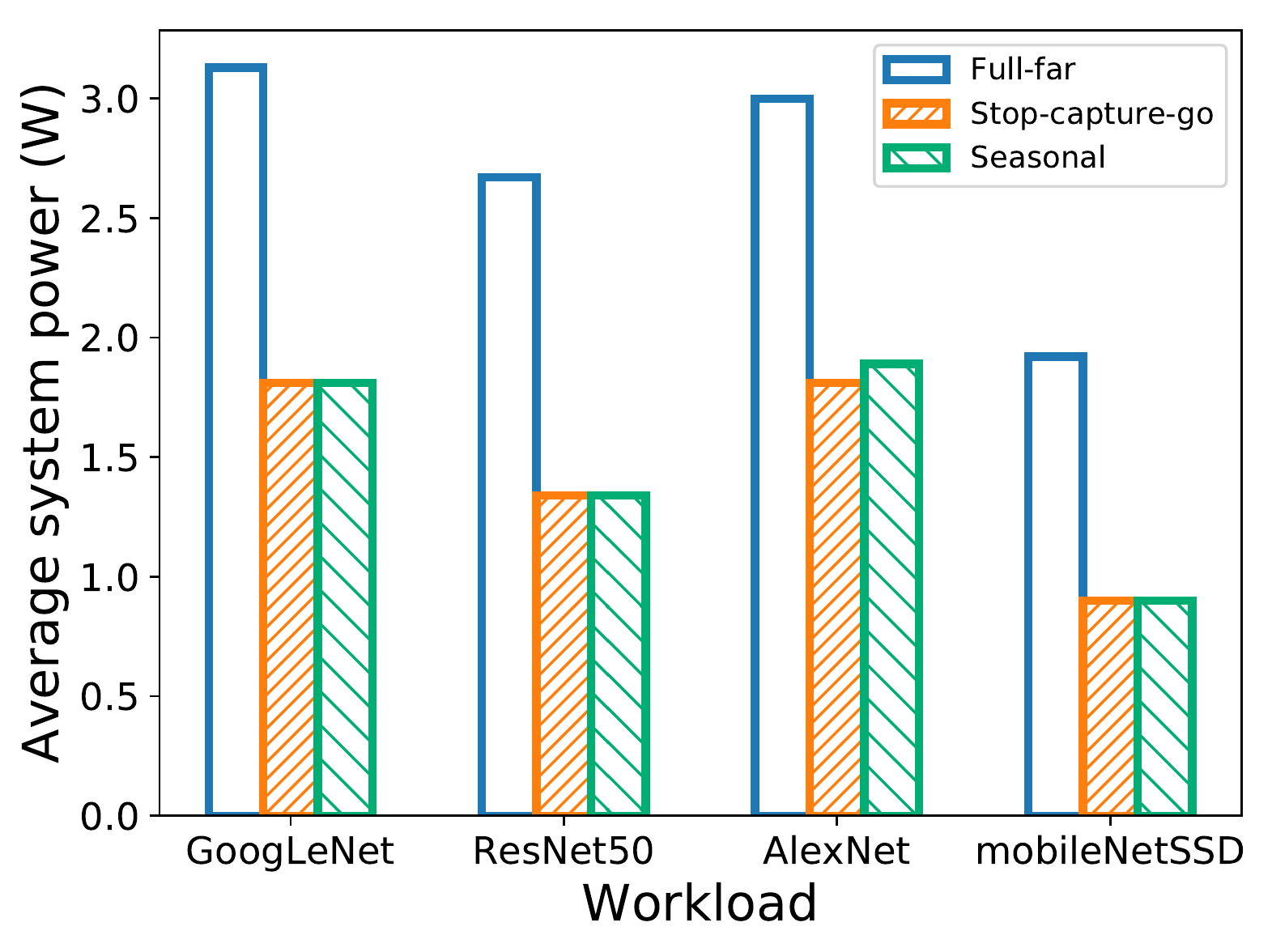}
    \caption{Don't care}
   \end{subfigure}
   \caption{Average system power varies with fidelity needs.
   For seasonal migration, raised duty cycles decrease system power due to more near-sensor operation.
   For stop-capture-go, raised duty cycles increase the system power due to more VPU sleep time, although improving performance by minimizing frame drops.
   }
   \label{fig:powerVsAppVsSNR}
   \vspace{-2em}
\end{figure*}

\subsection{Duty cycle}
Stagioni determines optimal duty cycles, based on power profile and imaging fidelity requirements.
Fig.~\ref{fig:dutyCycleVsappsVsSNR} shows a sensitivity analysis of duty cycle for a range of power dissipations across different fidelity needs.
This range of power profiles can represent different executions on Myriad2-like architecture or on ASIC, GPU, and FPGA architectures with different performance expectations.
We can see that the duty cycle varies widely, due to the strong interplay between the fidelity and the power profile.
While the application's power profile determines the steady-state temperature, the fidelity requirement determines the placement of thermal boundaries in the temperature trace.
This can result in a broad range of duty cycle based on where the thermal boundaries are situated in the temperature response, which we explain below.

If thermal boundaries are placed above the temperature response, Stagioni operates at 100\% duty cycle, i.e., in \NSP mode all the time.
This is relevant the steady-state temperatures of an application power profile are below thermal limits, e.g., $<$ \SI{1}{\watt}.
On the other hand, the boundary placement within the gradual rise and steeper fall region of the temperature trace means that the system spends more time in \NSP mode than \CAP mode, resulting in duty cycles greater than 50\%.
If the boundaries lie in the steeper rise and gradual fall region, this time, system spends more time in \CAP mode than \NSP mode, thereby leading to duty cycles $<$50\%.

\subsection{System Power Consumption}
Here, we examine system power during emulated workloads.
We find that stop-capture-go and seasonal migration substantially reduce system power compared to the status quo.
Fig.~\ref{fig:powerVsAppVsSNR} shows the system power for different applications for different policies, across different fidelity needs.
We see that stop-capture-go consumes the lowest amount of power among all the policies.
This is because stop-capture-go operates entirely on the near-sensor VPU for whole program execution in both \NSP and \CAP modes.
In contrast, seasonal migration operates on far-sensor VPU during \CAP mode and on near-sensor VPU during \NSP mode.
Thus, it consumes more power than stop-capture-go but less than full-far policy.

System power changes with fidelity demands due to change in duty cycle;
high fidelity pulls down the duty cycle, reducing efficiency.
This is evident in seasonal migration; we see higher power for higher app fidelities (higher SNR).
For stop-capture-go, a lower duty cycle increases VPU sleep time, while dropping frames from processing.
Therefore, we see power \emph{decrease} as we go from low to high app fidelity.
For full-far policy, there is no change in system power, as it doesn't create fidelity issues.

\subsection{Overhead}
We discuss policy execution overhead for seasonal migration and stop-capture-go policies.
As the system executes seasonal migration, it switches between near-sensor and far-sensor VPUs.
However, through the use of practically available techniques, task offload overhead can be kept to a minimum.
We use one such technique called pre-copy migration which pre-emptively transfers the state before the migration deadline.
Consequentially, one needs to only take care of synchronization between the VPU and SoC, which involves only basic handshaking operations incurring minimal overhead.
We measure this switching overhead on our emulation setup to be 100 $\mu$s, which is negligible in the context of the frame capture time, i.e., 33 ms.
Ergo, seasonal migration has no effect on the performance of the vision application.

For stop-capture-go, stop time determines the number of frame drops which could potentially lead to performance hit.
The actual performance loss depends on the duty cycle and effective frame rate when the system executes stop-capture-go will be scaled by a factor of the duty cycle which has interesting implications.
If the duty cycle is very high, then the performance loss would be minimal, e.g., 30 fps with 98\% duty cycle leads to an effective 29.4 fps.
On the other hand, a lower duty cycle can lead to a substantial performance loss, e.g., 30 fps with 40\% duty cycle leads to 12 fps, which is a reduction by more than 50\%.
Therefore, even though stop-capture-go consumes the lowest system power, it can hit the performance of the vision application when near sensor power and/or fidelity requirements are high.

\begin{figure}[t]
   \centering
   \begin{subfigure}[t]{0.46\columnwidth}
    \includegraphics[width=.9\textwidth]{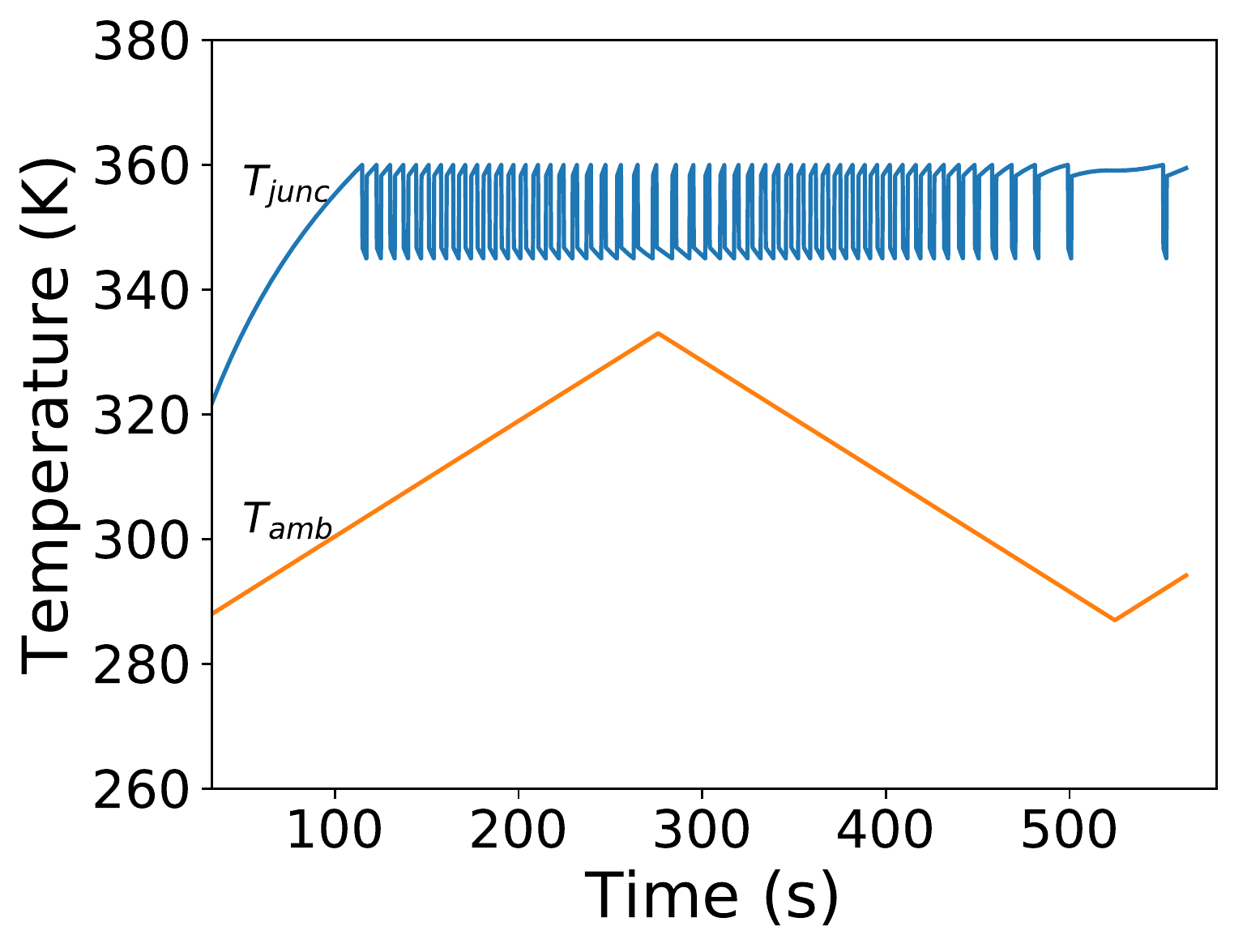}
    \caption{Adaptive to temperature.}
    \label{fig:tempTraceDynAmbiTemp}
   \end{subfigure}
   \quad
   \begin{subfigure}[t]{0.46\columnwidth}
    \includegraphics[width=.9\textwidth]{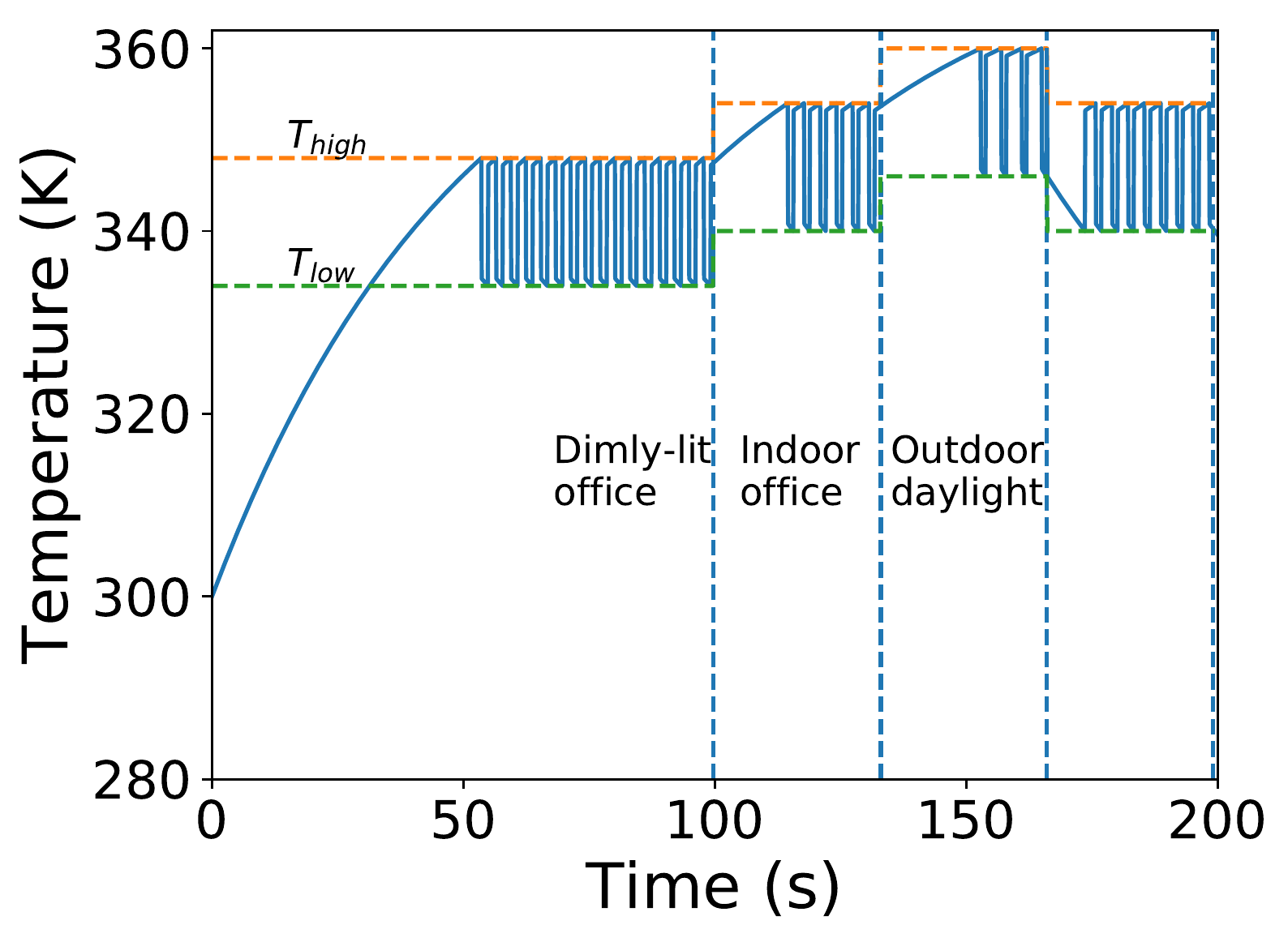}
    \caption{Adaptive to lighting.}
    \label{fig:tempTraceDynAmbiLight}
   \end{subfigure}
   \caption{Increasing ambient temperature (left) and/or decreasing ambient illumination (right) pulls $T_{steady}^{NSP}$ away from $T_{low}$ and pushes $T_{steady}^{CAP}$ close to $T_{high}$.
   Stagioni shifts thermal boundaries to smoothly adapt to different ambient conditions.}
   \vspace{-1em}
\end{figure}

\subsection{Situational awareness}
One feature of Stagioni that differs from traditional DTM techniques is situational awareness to dynamic ambient settings.
We find that Stagioni smoothly adapts thermal boundaries to match ambient temperature and lighting situations.

\noindent\textbf{Ambient temperature awareness:}~
Ambient temperature determines steady-state temperatures, which determine the warming and cooling times.
Higher ambient temperatures push $T_{steady}^{NSP}$ far from $T_{low}$ and push $T_{steady}^{CAP}$ close to $T_{high}$.
This forces the warming phase to take a steeper rise and the cooling phase to take a gradual fall in the exponential curve.
Thus, increasing ambient temperature decreases duty cycle and vice-versa.
We simulate the change in ambient temperature in our emulation platform, shown in Fig.~\ref{fig:tempTraceDynAmbiTemp}. Decreasing ambient temperature increases rise times and reduces fall times in the simulated temperature trace.
We also notice that Stagioni smoothly adjusts to the changes in ambient temperature.

\noindent\textbf{Ambient light awareness: }~
Lighting dictates fidelity requirements, changing $T_{high}$ and $T_{low}$.
Stagioni adapts to these changes.
We simulate change in illumination to generate a trace with random juggling between lighting scenarios.
We provide this trace as input to our runtime and collect the temperature trace.
Fig.~\ref{fig:tempTraceDynAmbiLight} shows the temperature trace overlaid with $T_{high}$ and $T_{low}$.
We can observe the smooth variation of temperature with light intensity.

\section{Conclusion}
Near-sensor processing can unlock energy-efficient imaging and vision, as demonstrated by recent academic and industrial efforts. 
However, we show that doing so hampers sensor fidelity due to thermal noise, thereby limiting the adoption of near-sensor processing.
Our characterization reveals that immediate drop in temperature can be realized within a short duration.
We use this observation to design principles for managing sensor temperature for efficient temperature regulation and high fidelity temperatures, while optimizing for system power.
To implement the policies, we design and implement the Stagioni runtime to manage sensor temperature, while fulfilling imaging needs.
Our work is the first runtime solution for stacked sensor thermal management.
We foresee our work as early steps to imaging-aware DTM techniques.

\section*{Acknowledgements}
The authors thank Linda Nguyen for her help with the figures and Microsemi for their generous support through SmartFusion2 hardware kit and software licenses.
This material is based upon work supported by the National Science Foundation under Grant No.1657602.

\bibliographystyle{ieeetr}
\bibliography{bib/intro,bib/bg,bib/char,bib/design,bib/impl,bib/eval,bib/related,bib/disc}

\begin{IEEEbiography}[{\includegraphics[width=1in,height=1.25in,clip,keepaspectratio]{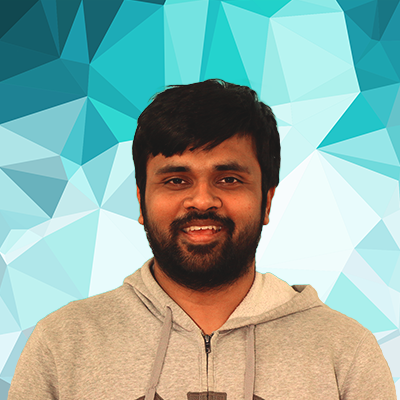}}]{Venkatesh Kodukula}
received his B.Tech degree in Electronics and Communications Engineering from National Institute of Technology, Warangal, India in 2011, and a MS degree in Computer Engineering from Arizona State University in 2018.
Currently, he is pursuing a Ph.D. degree in Computer Engineering from Arizona State, as well.
His research interests include runtime thermal management of sensors and hardware-software co-design for efficient computer vision.
\end{IEEEbiography}

\begin{IEEEbiography}
[{\includegraphics[width=1in,height=1.25in,clip,keepaspectratio]{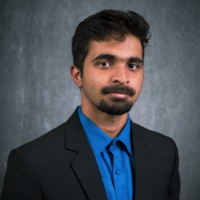}}]{Saad Katrawala}
received his BE degree in Electronics and Instrumentation from Birla Institute of Technology and Science, Pilani, India in 2017.
He is currently pursuing his MS in Computer Engineering from Arizona State University, Tempe.
His research interests include the areas of operating systems and computer architecture.
\end{IEEEbiography}

\begin{IEEEbiography}[{\includegraphics[width=1in,height=1.25in,clip,keepaspectratio]{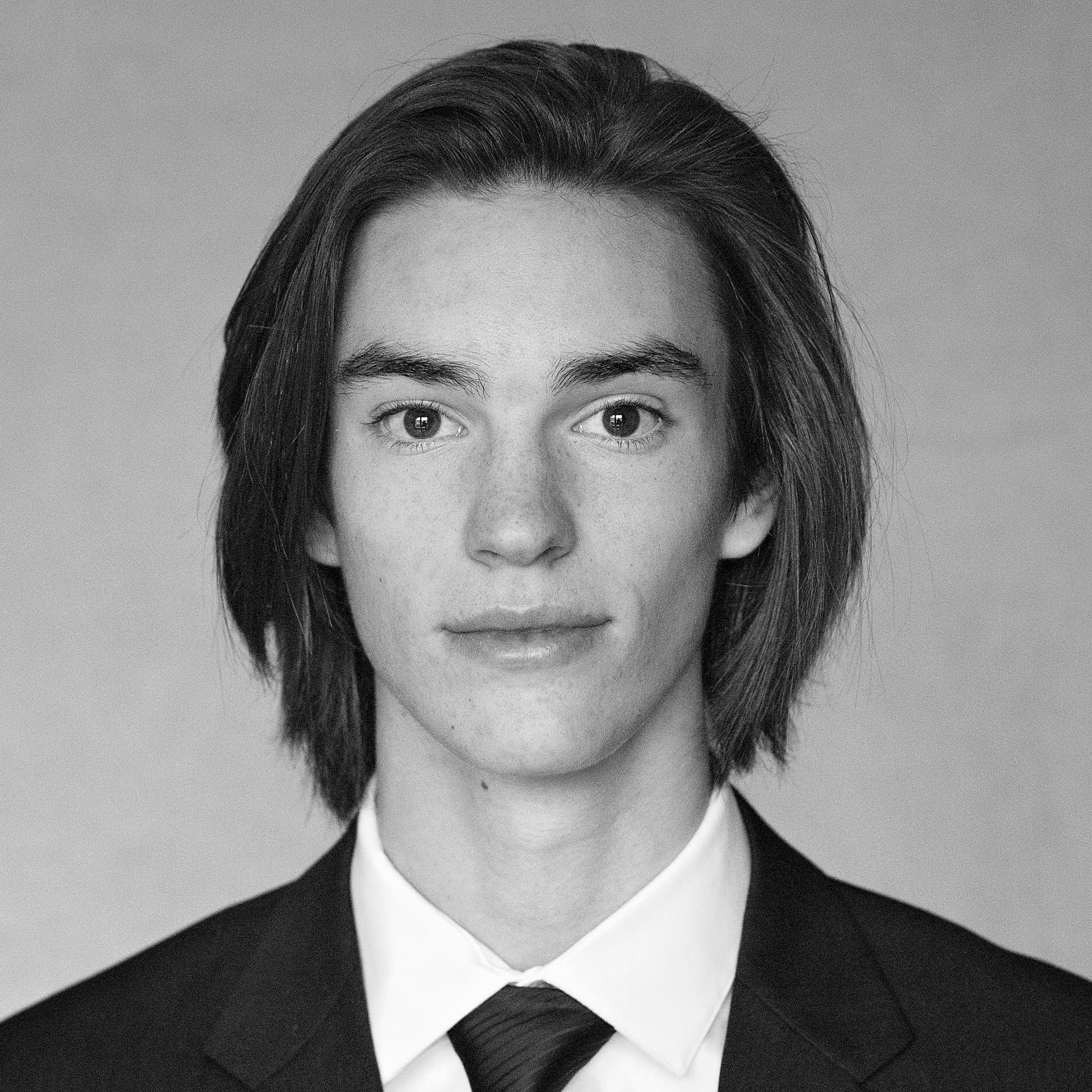}}]{Britton Jones}
is currently receiving the BSE degree in Electrical Engineering at Arizona State University.
Following that, he hopes to receive the MS degree in Electrical Engineering from Arizona State, as well.
He is a third-year student and has been involved in academic research for two years on image processing and computer vision.
\end{IEEEbiography}

\begin{IEEEbiography}[{\includegraphics[width=1in,height=1.25in,clip,keepaspectratio]{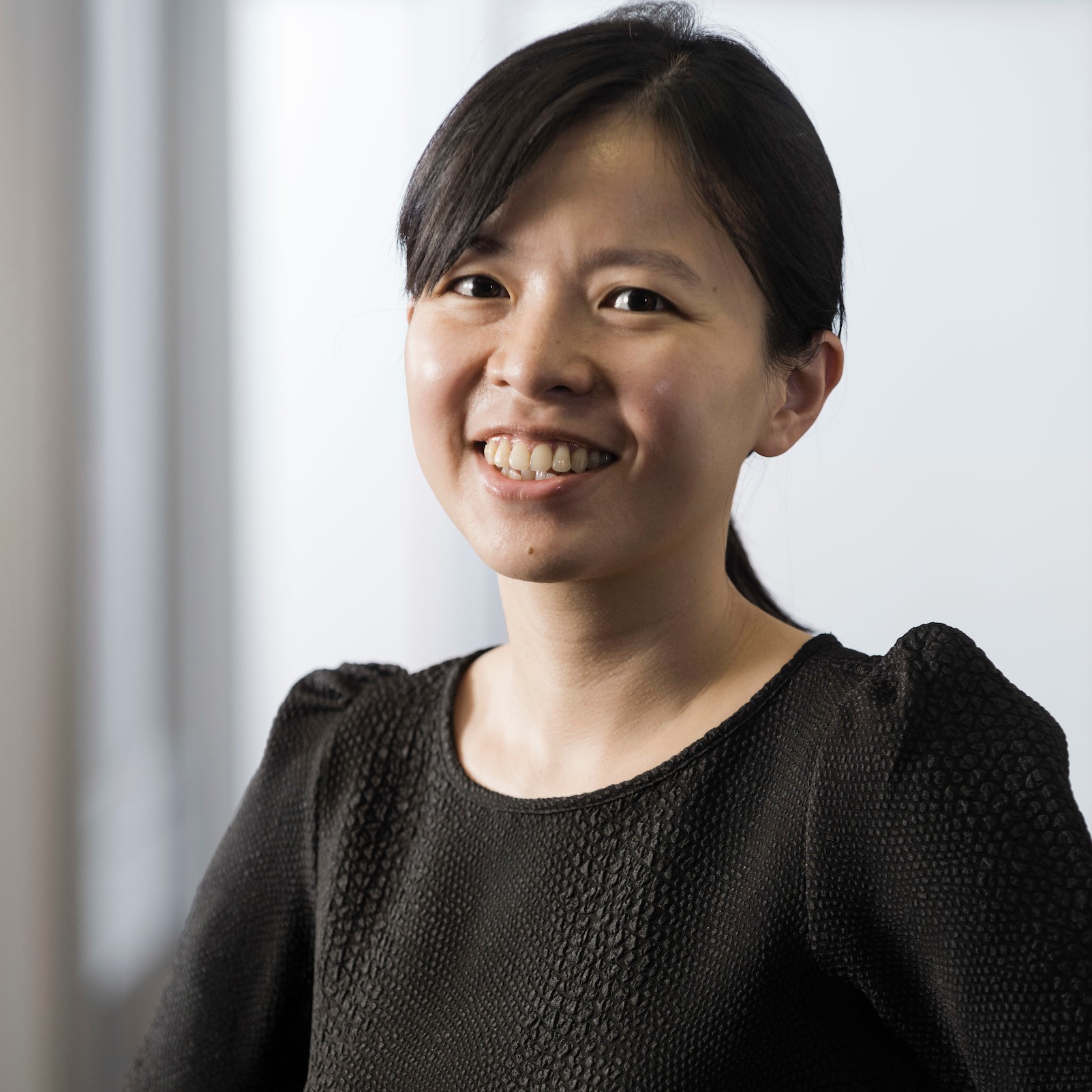}}]{Carole-Jean Wu}
received her BS degree in Electrical and Computer Engineering from Cornell University (2006) and the MA and the PhD degrees (2008, 2012) in Electrical Engineering from Princeton University.
She is currently an Associate Professor with the School of Computing, Informatics, and Decision Systems Engineering at Arizona State University.
 Her research interests include high-performance and energy-efficient computer architecture through hardware heterogeneity, energy harvesting techniques, temperature and energy management.
More recently, her research has pivoted into designing systems for machine learning.
She is a senior member of both ACM and IEEE.
\end{IEEEbiography}

\begin{IEEEbiography}[{\includegraphics[width=1in,height=1.25in,clip,keepaspectratio]{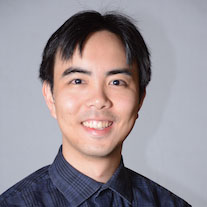}}]{Robert LiKamWa}
received his B.S. (2010), M.S. (2012), and Ph.D. (2016) degrees in Electrical and Computer Engineering from Rice University. He is currently an assistant professor in the School of Arts, Media and Engineering and the School of Electrical, Computer, and Energy Engineering.
His research interests include operating system designs for energy-efficient visual sensing and mobile systems for augmented reality and virtual reality.
\end{IEEEbiography}

\end{document}